\documentstyle[epsfig]{mn}
\newcommand{\iue}{{\it IUE}}
\newcommand{\fuse}{{\it FUSE}}
\newcommand{\hst}{{\it HST}}
\newcommand{\msec}{\mbox{$\mu{\rm s}\;$}}
\newcommand{\kms}{\mbox{\thinspace km\thinspace s$^{-1}$}}
\newcommand{\msunyr}{\mbox{\thinspace M$_\odot$\thinspace yr$^{-1}\;$}}
\newcommand{\msun}{\mbox{\thinspace M$_{\odot}$}}

\newcommand{\mdot}{\mbox{$\stackrel{.}{\textstyle M}$}}
\newcommand{\lya}{\mbox{Ly$\,\alpha$}}
\newcommand{\IV}{\mbox{\thinspace{\sc iv}}}
\newcommand{\II}{\mbox{\thinspace{\sc ii}}}
\newcommand{\III}{\mbox{\thinspace{\sc iii}}}

\newcommand{\V}{\mbox{\thinspace{\sc v}}}
\newcommand{\VI}{\mbox{\thinspace{\sc vi}}}
\newcommand{\IVl}{\mbox{\thinspace{\sc iv}}$\,\lambda\,$}
\newcommand{\IIl}{\mbox{\thinspace{\sc ii}}$\,\lambda\,$}
\newcommand{\IIIl}{\mbox{\thinspace{\sc iii}}$\,\lambda\,$}
\newcommand{\Il}{\mbox{\thinspace{\sc i}}$\,\lambda\,$}
\newcommand{\Vl}{\mbox{\thinspace{\sc v}}$\,\lambda\,$}
\newcommand{\VIl}{\mbox{\thinspace{\sc vi}}$\,\lambda\,$}
\newcommand{\lbol}{\mbox{$L_{\rm{bol}}$}}

\title[Time-resolved UV spectroscopy of IX Vel and V3885 Sgr] {Testing
the line-driven disk wind model: time-resolved UV spectroscopy of
IX~Vel and V3885~Sgr}
\author[L. E. Hartley, J. E. Drew, K. S. Long, C. Knigge, D. Proga]{L. E. Hartley$^1$,
J. E. Drew$^1$, K. S. Long$^2$, C. Knigge$^3$, D. Proga$^{4,5}$\\
$^1$Department of Physics, Blackett Laboratory, ICSTM,
Prince Consort Road, London SW7 2BW\\
$^2$Space Telescope Science Institute, 3700 San Martin Drive, Baltimore, MD 21218\\
$^3$Department of Physics \& Astronomy, University of Southampton,
Southampton, SO17 1BJ\\
$^4$LHEA, Code 662, 
NASA Goddard Space Flight Center, Greenbelt, MD 20771\\
$^5$JILA, University of Colorado, Boulder, CO 80309-0440}

\begin{document}
\maketitle

\begin{abstract}
   To confront the predictions of the most recent line-driven disk wind models
with observational evidence, we have obtained \hst\ STIS (1180--1700\thinspace\AA) echelle
spectra of the nova-like variables IX~Vel and V3885~Sgr at three epochs.  The
targets were observed in timetag mode for $\sim$2000\thinspace sec on each occasion,
allowing us to study the spectral time evolution on timescales down to
$\sim$10\thinspace sec.  The mean UV spectra are characterised by the wind signature of
broad blueshifted absorption in \lya, N\Vl1240,
Si\IVl1398, C\Vl1549 and
He\IIl1640. There is very little redshifted emission other
than in C\IV.  Narrow blueshifted absorption dips, superposed on the
broad absorption at around $-900$\kms, accompany periods of
well-developed wind activity.  The continuum level and mean line profiles
vary markedly from observation to observation -- with the wind signatures
almost disappearing in one epoch of observation of IX~Vel.  The strong
positive correlation between UV brightness and wind activity predicted
by line-driven disk wind models is disobeyed by both binaries.

   The wind signatures in IX~Vel's UV spectrum are revealed to be remarkably
steady on timescales ranging from $\sim$10 to $\sim$1000\thinspace sec.  More
variability is seen in V3885~Sgr, the binary with the lower opacity outflow.
But there is only one epoch in which the line profile changes
significantly in $\sim$100\thinspace sec or less.  Narrow absorption dips, when present, show only
smooth, small changes in velocity.  We surmise these may trace the white
dwarf's orbital motion.  The near-absence of line profile variability on
the shorter 10- to 100-sec timescales, and the lack of correlation between
wind activity and luminosity, could both arise if a non-radiative factor
such as the magnetic field geometry controls the mass loss rate in these
binaries.
\end{abstract}


\section{Introduction}

Accretion disks are present in many astrophysical systems, with firm
evidence for their existence in environments as diverse as those of young
stars \cite{98hartmann}, close binary systems and AGN (Frank, King \& Raine
1985).  Objects known to include hot optically-thick disks typically provide
clear spectroscopic evidence of mass loss in the form of broad blueshifted UV
absorption features.

Currently there are two main driving mechanisms envisaged for disk winds. The
more widely studied of these, regarded as the only option for low-mass
YSOs, relies on MHD processes (e.g. Ouyed \& Pudritz 1997,
Hirose et al. 1997, Konigl \& Pudritz 2000).
In high luminosity systems an alternative or additional method for
driving an outflow is radiation pressure mediated by line opacity (hence
the term `line-driven wind'). Until recently the multi-dimensional modelling
required to deal with the geometry of a line-driven disk wind was beyond
reach.  However, this problem has now been approached
numerically, with greater success, in the models of Pereyra, Kallman \&
Blondin (1997) and Proga, Stone \& Drew (1998, hereafter PSD).  As a result
of this work, the radiation-driven wind model already provides some clear-cut
predictions that can be put to the test against observations.  Pre-eminent
among these is the expectation that the wind mass loss rate will show a strong
dependence on system luminosity (see Drew \& Proga 2000).

PSD also predict unstable, clumpy outflow in the
regime where the accretion disk, rather than the central object, contributes
most of the radiant luminosity.  Indeed, PSD's simulations suggested that the
dominant mass loss stream flowing away from the inner disk will include
denser clumps of gas that fail to reach escape velocity and collapse back
onto the disk at larger radii.  A better treatment of the line-driving
(Proga, Stone \& Drew 1999) preserves this impression of a highly structured, clumpy
outflow.  When observed spectroscopically with a time resolution of less than
the typical flow timescale this phenomenon could make itself apparent as
narrow absorption components, superposed on the broad blueshifted absorption
profile, sweeping both away from and in towards line centre.

With the aim of looking for the spectral signatures of disk-wind variability
on suitably short timescales we have obtained {\it Hubble Space Telescope} (\hst)
observations of two high-state cataclysmic variables, using the Space
Telescope Imaging Spectrograph (STIS). Specifically, our targets are the
nova-like variables: IX Vel and V3885 Sgr.  These
binaries provide examples of disk winds where the line-driving mechanism may
dominate over MHD processes in powering mass loss.  Both IX~Vel and V3885~Sgr
are made up of a low mass star overflowing its Roche lobe and transferring
matter to a non- or weakly-magnetic white dwarf (WD) via an accretion disk (i.e.
they are UX~UMa type cataclysmic variables).  Both are known to persist in a
high brightness state, powered by high mass accretion rates
($\la10^{-8}$\msunyr).  This puts these objects just within the domain
where disk luminosities are $\sim0.001L_{\rm{E}}$ (where $L_{\rm{E}}$ is the Eddington
luminosity) and spectral line opacity may be sufficient to power significant
mass loss \cite{2000drew}.  Essentially all of the light from these systems at
optical and ultraviolet wavelengths is due to accretion.  A further
factor influencing target choice is that IX~Vel and V3885~Sgr are the
brightest of the known UX~UMa binaries.
With the possibility of high-quality observation of the wind-shaped UV line
profiles, these binaries can provide a particularly good test bed for
line-driven disk wind models.

On timescales of weeks and months, IX~Vel and
V3885~Sgr vary by several tenths of a magnitude.  In order to allow this
secular variation to provide a sampling of a modest range of luminosity states
, both targets were observed three times each at intervals of at least a month.
Both targets are non-eclipsing systems. The broad blueshifted
absorption features seen in the UV, that are indeed a common feature of the
lower inclination UX UMa systems, are not usually very deep, making them
ideal for studying the fine structure of wind-formed lines. To obtain the
high time and wavelength resolution required for this programme, STIS
was employed in its time-tag observing mode, using an echelle to disperse
the light. This allows the observer to select the desired wavelength or time
resolution at the point of data extraction and calibration.  To ensure a
minimum S/N ratio there is necessarily a trade-off such that higher time
resolution will result in a reduced useable spectral resolution and
vice-versa.

This is not the first use of the \hst\ spectrographs for high time
resolution studies of cataclysmic variable disk winds: Prinja et
al. (2000a), (2000b) report results of Goddard High Resolution
Spectrometer (GHRS) grating observations of respectively BZ~Cam and
V603~Aql.  These data gave a time resolution of $\sim$80 and $\sim$30
seconds, timescales that are also comparable with the likely outflow
timescales in these binaries (of order a few tens up to $\sim$100
seconds).  Significant absorption line profile variability on
timescales down to $\sim60$~sec was reported.  The key differences
between these published studies and the present study lie in data
quality -- our brighter targets allow the use of the higher resolution
echelle -- and in the prior knowledge of the targets' UV spectral
characteristics.

The organisation of this paper is as follows.  First, we present a
brief description of the STIS/time-tag observations (section
\ref{sec:obs}).  The results are then presented in two ways: in
section \ref{sec:secular} we describe the time-averaged spectrum
obtained at each epoch of observation, and then in \ref{sec:timevar}
we re-present the data as spectral time-series at 30-sec time-resolution.
In section \ref{sec:discus}, the results obtained are put into context
and we review the variability uncovered in IX~Vel and V3885~Sgr and
compare with the prior examples of BZ~Cam and V603~Aql.  Prompted by
our finding that wind activity seems to show no correlation with
luminosity (thereby challenging the pure radiation-driven disk wind
model), we revisit earlier-generation {\it International Ultraviolet
Explorer} (\iue) spectroscopy of IX~Vel in order to seek further
evidence of this (section \ref{subsec:discus3}).  We close with some
thoughts on the next move in the quest for a fuller understanding of
the driving of these accretion disk winds.

\section{Observations and Data Extraction}
\label{sec:obs}

Observations were performed by the STIS instrument on the Hubble Space
Telescope, with the far ultraviolet (FUV) MAMA detector
(1140--1735\thinspace\AA). The E140M echelle grating with a central
wavelength of 1425\thinspace\AA ~was used. This configuration provides a
resolution of $\sim10\kms$. V3885 Sgr was observed with the
0.2\arcsec$\times$0.2\arcsec~aperture. IX Vel, the brighter target,
had to be observed with the 0.1\arcsec$\times$0.03\arcsec ~slit to
reduce the likelihood of buffer overflow. Each
observation was performed in a single telescope orbit for the maximum
available time, with the detector set to time-tag mode. The dates and
exposure times of our observations are given in table
\ref{tab:obs_dates}.

\begin{table}
\caption{The observation dates and times for each \hst\ observation.}
\label{tab:obs_dates}
\begin{tabular}{@{}lcccc}
Object	&Date	&Label	&U.T. start time&Exposure	\\
	&(2000)	&	&		&(s)		\\
IX Vel	&3rd April&I1	&19:37:06	&1750		\\
	&30th May&I2	&12:57:46	&1750		\\
	&19th Aug&I3	&08:31:29	&1750		\\
V3885 Sgr&30th April&V1	&08:01:20	&2480		\\
	&20th Aug&V2	&03:09:13	&2480		\\
	&13th Nov&V3    	&01:43:02	&2480		\\
\end{tabular}
\end{table}

All data calibration was performed with {\sc IRAF} software, using the
{\sc STSDAS} package produced by the Space Telescope Science Institute
(STScI). Time-tag is a photon counting mode, which provides an events
stream with 125-\msec time-resolution.  This can be integrated over any
selected time resolution (time bin) to produce a set of raw images of
the echelle output, from which a 1-D spectrum is extracted and
calibrated. The calibration, including doppler correction (on to the
heliocentric frame) for the telescope's motion, was performed with
STScI's {\sc CALSTIS} package, using the suggested best reference
files.

\section{Secular changes in the mean UV spectrum}
\label{sec:secular}

\subsection{IX Vel}
\label{subsec:ixvel1}

\begin{table}
\caption{Phase and velocity data for IX Vel, relative to the
spectroscopic ephemeris and using the data of Beuermann \& Thomas
(1990). Phase and velocity are given for the mid-point of the
observation.}
\label{tab:ix_phase}
\begin{tabular}{@{}lccc}
Period (d) &&\multicolumn{2}{l}{$0.1939\pm0.0020$} \\ $K$-velocity
(\kms) &&\multicolumn{2}{l}{$138\pm5$} \\ $\gamma$-velocity (\kms)
&&\multicolumn{2}{l}{48} \\
\hline
Observation date 	&3rd April 	&30th May 	&19th August \\
\hline
Dataset 		&I1 		&I2 		&I3
\\ Phase 		&$0.819$ 	&$0.226$ 	&$0.237$
\\ Velocity (\kms) 	&$173\pm5$ 	&$-88\pm5$ 	&$-90\pm5$ \\
\end{tabular}
\end{table}

IX Vel is the brightest known nova-like variable, with apparent
magnitude given at different times as $m_{\rm{v}} = 9.4$
\cite{77klare} and $m_{\rm{v}}=9.8$
\cite{83schild}. It has a period of 4\fh65 and an orbital inclination
estimated to be $60^\circ\pm5^\circ$ \cite{90beuermann}. It exhibits
flickering in the optical of $\sim0.1$ mag on a timescale of minutes
\cite{84garrison}.

Studies of the wind-formed UV resonance line profiles of IX Vel, based on 
high-resolution \iue\ spectra, have previously been carried out by 
Mauche~\shortcite{91mauche} and by Prinja \& Rosen~\shortcite{95prinja}. 
Mauche (1991) found subtle variations in the wind-formed UV line profiles, 
which he was able to link to orbital phase and he noted the presence
of superposed narrow absorption components. Prinja \& 
Rosen~\shortcite{95prinja} comment on the narrow absorption components
as well and compare IX~Vel with other high-state non-magnetic CVs.

\begin{figure*}
\vspace*{10cm}
\includegraphics{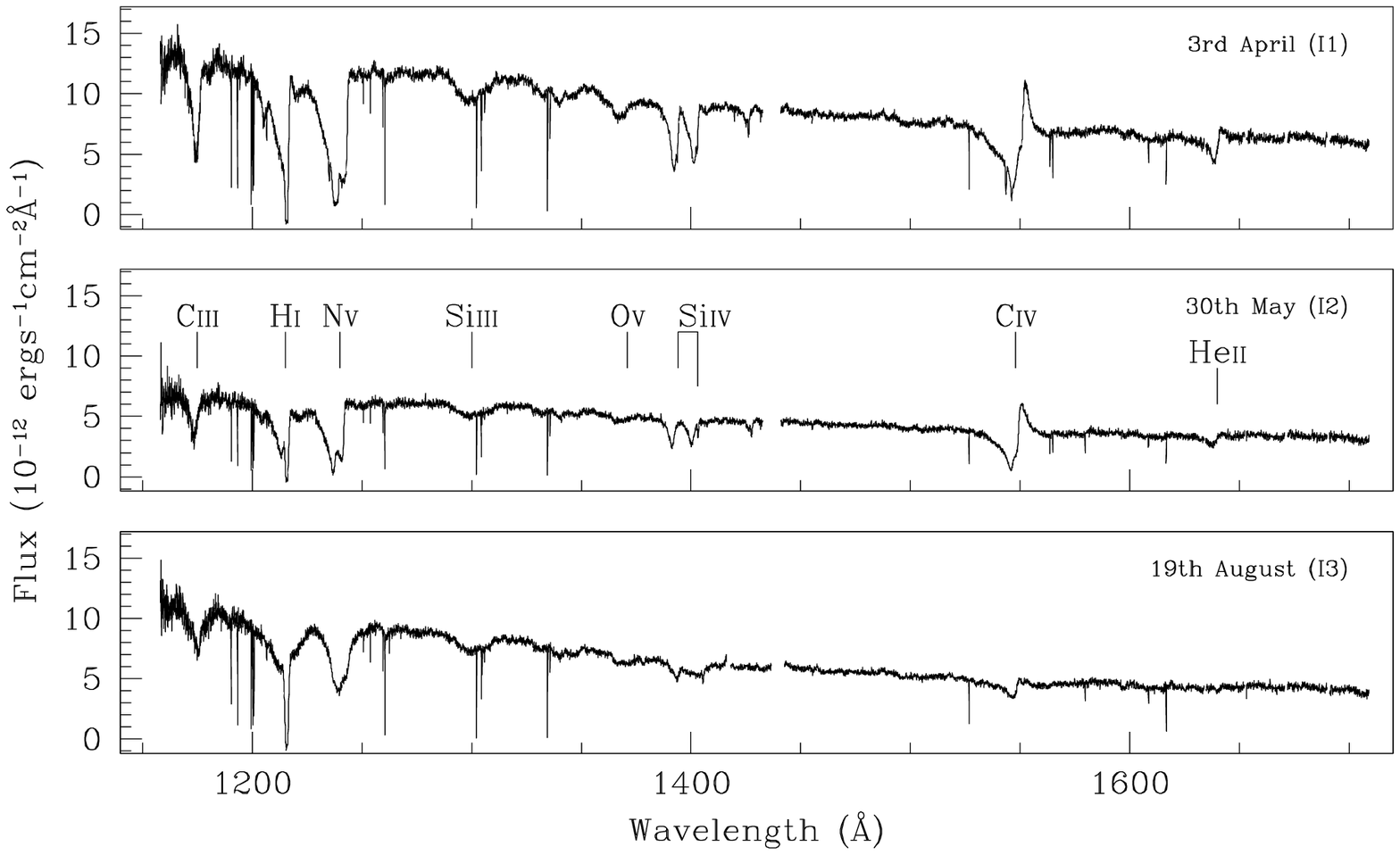}
\caption{The summed spectrum for the (from top to bottom) 3rd
April, 30th May and 19th August observations of IX Vel. These are
respectively referred to as I1, I2 and I3 in the text and in tables.
Stronger line features are identified in the middle panel.}
\label{fig:ixvel}
\end{figure*}

Figure \ref{fig:ixvel} shows the mean spectrum obtained at each visit to the 
target (hereafter referred to as I1, I2 and I3). In the first two datasets 
IX~Vel's spectrum is characterised by broad blueshifted absorption profiles 
in \lya, N\Vl1240, Si\IVl1398, C\IVl1549~and 
He\IIl1640. Redshifted emission is also present at low contrast 
in \lya\ and He\II, and much more prominently in C\IV.  However, in the third 
spectrum, only C\IV~retains a weak P$\:$Cygni profile, while the profiles of 
the other lines are more symmetrical around their rest wavelengths.

To ease comparison between observations obtained at different times,
we have applied radial velocity corrections that should place the data
in the WD rest frame.  The most up-to-date spectroscopic ephemeris is
given by Beuermann \& Thomas (1990).  These authors also calculated a
mean disk $K$-velocity of $138\pm5\kms$.  We have used these data to
estimate the expected mean WD radial velocities at the times of the
\hst\ observations. These are given in table \ref{tab:ix_phase}.

Our calculations suggest that I2 and  I3 were recorded when the target
was at almost the same position in its orbit and when the WD line of
sight (LOS) velocity
was  approaching a maximum towards  the observer.   I1 data were taken
when the WD's line  of sight velocity  was directed away  from the
observer.  We cross-correlated I1 and I2 to determine
the    velocity  shift between the redward absorption edges of the
N\Vl1240, Si\IVl1398\ and C\IVl1549\ lines in
each dataset. This edge is redshifted in I2 when compared to I1, as
expected from the predicted orbital motion.  However, we measured a
velocity difference   between  the two  datasets     of
$400\pm50\kms$ -- apparently too large to  be solely  due to the  orbital
motion.  This should be $261\pm7$\kms,  if the
Beuermann \& Thomas   (1990)  ephemeris predicts the   correct phases.
Furthermore, we  note  that,  according to the    optically determined
$K$-velocity semi-amplitude,  the  measured velocity difference should
not exceed $276\pm10$\kms .

\begin{table*}
\begin{minipage}{12.2cm}
\caption{Measurements of continua and wind lines in IX Vel.  `Flux' refers to the mean
continuum flux level in the range 1260--1270\thinspace\AA. `Index' refers to the index
of a power law continuum fit to the data. The quantities EW and
$v^{\rm abs}_{\rm max}$ both depend on eye estimates of the background continuum
level. They are respectively, the equivalent width of the absorption component
and the maximum velocity of blueshifted absorption. The EW measurement
of \lya\ includes the interstellar component in every case. Where `$s$'
appears after an EW measurement, the measured profile is symmetric,
rather than obviously dominated by blueshifted absorption. A colon after a velocity indicates
that the measurement is highly uncertain. The typical error on the equivalent
width measurements is 6 per cent.}
\label{tab:ixvel_vels}
\begin{tabular}{@{}l|cc|cc|cc}
&\multicolumn{2}{c}{3rd April}	&\multicolumn{2}{c}{30th
May}&\multicolumn{2}{c}{19th August}\\
&\multicolumn{2}{c}{I1}&\multicolumn{2}{c}{I2}&\multicolumn{2}{c}{I3}\\
\hline
Flux&\multicolumn{2}{c}{$11.5\pm0.5\times10^{-12}$}&\multicolumn{2}{c}
{$6.0\pm0.3\times10^{-12}$}&\multicolumn{2}{c}{$8.9\pm0.4\times10^{-12}$}
\\ Index &\multicolumn{2}{c}{$-2.27\pm0.10$}
&\multicolumn{2}{c}{$-2.20\pm0.10$}
&\multicolumn{2}{c}{$-2.70\pm0.10$} \\
\hline
Lines:		&EW (\AA)	&$v^{\rm{abs}}_{\rm{max}}$	&EW (\AA)	&$v^{\rm{abs}}_{\rm{max}}$	&EW (\AA)	&$v^{\rm{abs}}_{\rm{max}}$	\\
\hline
C\III$\lambda1175$	&$2.9$		&$-2300\pm200$	&$2.3$		&$-2000\pm200$	&$2.7^s$		&$-2000\pm200$	\\
\lya$\lambda1216$	&$5.4$		&$-4000:$	&$5.1$		&$-4000:$	&$6.3^s$		&$-4000:$	\\
N\V$\lambda1240$	&$9.1$		&$-3300\pm200$	&$7.6$		&$-2600\pm200$	&$5.6^s$		&$-2800\pm300$	\\
Si\IV$\lambda1398$&$4.9$		&$-2000\pm400$	&$3.7$		&$-1800\pm400$	&$3.0^s$		&$-1400\pm400$	\\
C\IV$\lambda1549$	&$7.1$		&$-5000:$	&$7.2$		&$-5000:$	&$2.4\:\:$		&$-5000:$	\\
He\II$\lambda1640$&$2.1$		&$-3000:$	&$1.8$		&$-3000:$	&$0.1^s$		&-		\\
\end{tabular}
\end{minipage}
\end{table*}

Figure 2 presents close-ups of the higher-contrast line profiles seen
in figure \ref{fig:ixvel}, with the C\IVl1549~line shown in more
detail in figure \ref{fig:ixvel_civ}.  Maximum blueshift velocities
measured from these are given in table \ref{tab:ixvel_vels}.

In I1 and I2 the strong lines possess similar
characteristics. \lya~displays an asymmetric profile with blueshifted
absorption. This is the first time that a significant \lya~wind
signature has been noted for this object.  We also detect redshifted
\lya~emission to about 1000\kms.  Other prominent lines, C\IIIl1176,
N\Vl1240~and Si\IVl1398~have a structured asymmetric profile, with
broad blueshifted absorption, but no emission. He\IIl1640~is also in
absorption to $\sim-3000$\kms, with an emission component extending to
$\sim1000$\kms.  The C\IVl1549~line is the broadest line, displaying
absorption to $\sim-5000\kms$ and emission extending redwards to at
least $1500\kms$ (see table \ref{tab:ixvel_vels}). This is consistent
with previous studies, e.g. Prinja \& Rosen (1995), which have found
that the C\IV~line tends to be formed over a larger volume than, for
example, Si\IV~and N\V.

The maximum velocity of blueshifted absorption is related to the
maximum wind velocity, so it is interesting to note that this value
varies little between I1 and I2 for all lines, despite the apparent
drop of nearly a factor 2 in the UV continuum flux between I1 and I2.
Also, the absorption minimum of all lines remains fairly constant at
around $-500$\kms, although the line profiles in I1 are more skewed.

We have measured absorption equivalent widths (EW) of the stronger
lines (see table~\ref{tab:ixvel_vels}).  Continuum placement was by
eye and we estimate the typical error in each EW determination is
$\sim$6~per cent.  There is a systematic drop in EW from I1 to I2 in
that the I2 EWs are $80\pm7$ per cent of their I1 values for the
C\III, N\V, Si\IV~and He\II\ transitions.  The uniformity of this
change indicates no clear shift either up or down in the wind's
ionization state. This, in turn, is an indication that the observed
line transitions arise in ions that are abundant, rather than trace,
ion stages for their respective elements.  If any of them were from
trace species we would expect them to stand out as markedly more
sensitive than the others (e.g. either the high-excitation He\IIl1640
or the lower ionization Si\IVl1398 or \lya, all likely to be
relatively low opacity line profiles, would show a bigger equivalent
width change). The C\IV\ EW shows essentially no change. This is not
surprising for what is often deduced to be a highly opaque transition.

In I3 a broader, more symmetric absorption profile, without redshifted
emission, is apparent in a range of transitions (for C\IIIl1176 the
profile spans\ $\sim\pm2000$\kms, \lya\ spans $\sim\pm3200$\kms , and
N\Vl1240 spans at least $\sim\pm2800$\kms~and perhaps as much as
$\sim\pm3100$\kms, given that \lya~obscures its blue wing).  Beuermann
and Thomas (1990) derive a WD mass of 0.8\thinspace($+0.16$,
$-0.11$)\msun .  Using this datum in the WD mass-radius relation given
by Nauenberg (1972), we derive a maximum line-of-sight Keplerian disk
velocity of about 3400\kms\ for $i=60^\circ$. This permits the \lya\
and N\V\ profiles seen in I3 to be rotationally broadened disk
photospheric features. In Si\IV, only the bluer component of the
doublet, which has the higher transition probability, is present and
it appears to sit on a broad, shallow absorption trough. The
C\IVl1549~P$\:$Cygni profile is uncharacteristically weak, showing
little emission, while He\IIl1640 has disappeared. This appears to be
a rare instance of an almost pure high-state disk line spectrum, all
but free of `contaminating' wind features. At the same time, it is
doubtful that the wind still exists at unchanged strength but with a
higher degree of ionization than witnessed in I1 and I2.  This is
because we would then expect either to continue to observe significant
blueshifted N\V\ absorption or to detect significant He\IIl1640
emission (due to recombination of He$^{2+}$ which would have to be
very abundant).

\begin{figure*}
\vbox to220mm{
\includegraphics{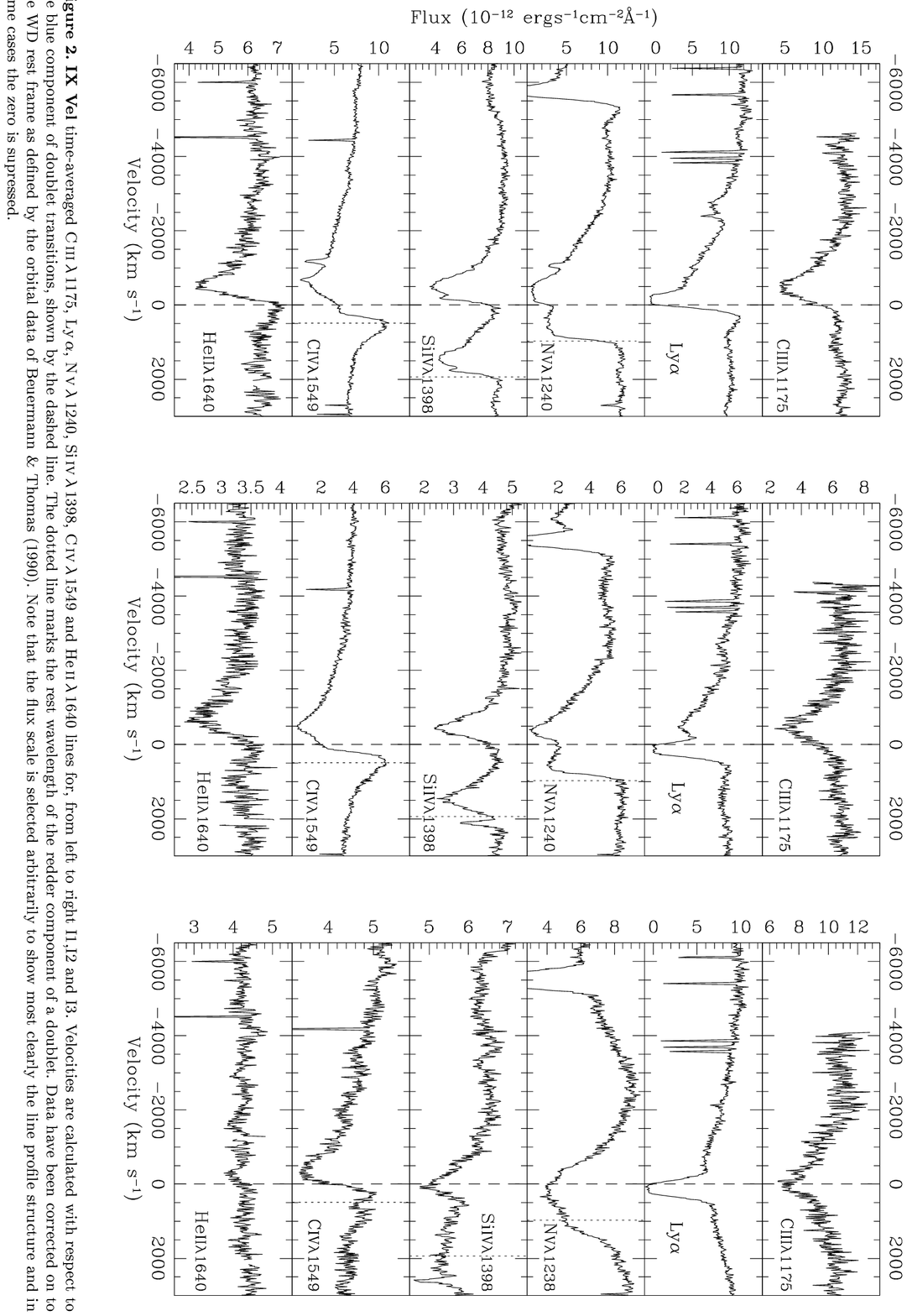}
}
\label{landfig1}
\end{figure*}
\setcounter{figure}{2}

A power law continuum fit was made to each mean spectrum, defining the
wavelength ranges 1260--1270\thinspace\AA, 1460--1470\thinspace\AA~and
1660--1670\thinspace\AA~as line-free continuum windows. Table
\ref{tab:ixvel_vels} lists the power law indices and the average flux in the 
range 1260--1270\thinspace\AA~for each dataset.

Both I1 and I2 fit to essentially the same power-law index
($F_\lambda\propto\lambda^{-2.2}$).  The continuum level of I3 lies in
between that of the other two datasets, but fits to a steeper power
law ($F_\lambda\propto\lambda^{-2.70}$), and hence is a `bluer'
spectrum.  The near-absence of wind signatures indicates a
greatly-reduced mass loss rate (by an order of magnitude or more).
These two attributes can be linked, since Balmer continuum emission
will be more pronounced in a denser wind.  The reddening of the
observed continuum in I1 and I2 then follows from the fact that Balmer
continuum emission on its own rises steeply with increasing wavelength
up to the Balmer limit at 3646\thinspace\AA . In this picture, I3 is
as blue a spectrum as it is because of the near absence of this redder
component.  The presence of Balmer continuum emission has also been
noted in the UV spectrum of UX UMa by Knigge et al. (1998) and
Baptista et al. (1998).

Finally, we draw attention to the surprising collapse of wind activity
in I3 at a time when the UV continuum was apparently at a level
intermediate between that seen in I1 and I2.  This spoils a clear
correlation between UV brightness and strength of wind signature and
raises a problem for the concept of a radiation-driven disk wind.  We
return to this in section \ref{subsec:discus3}.

\begin{figure}
\vspace{13cm}
\includegraphics{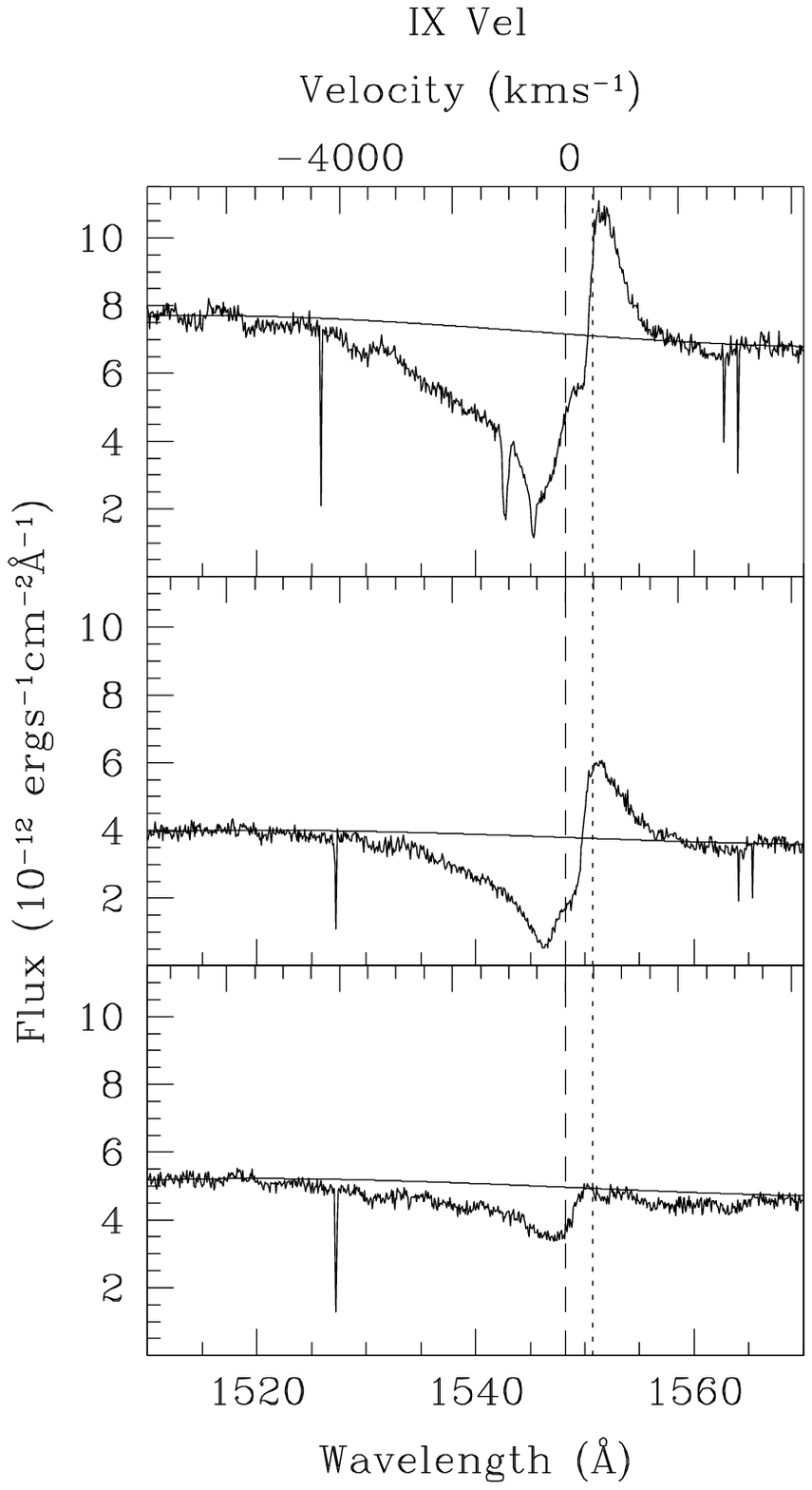}
\caption{Time-averaged C\IVl1549 for the IX~Vel observations I1, I2 
and I3, in order from top to bottom. Velocities (marked on the upper axes) 
are calculated with respect to the blue component (1548.2\thinspace\AA) of the doublet 
at rest, shown by the dashed vertical line. The dotted line marks the rest 
wavelength of the 1550.7\thinspace\AA ~component.  The smooth solid line is a 4th order 
fit to the continuum in the vicinity of the C\IV\ feature that is included to 
convey in impression of the full width of the profile. Data have been
corrected on to the WD rest frame as defined by the orbital data of
Beuermann \& Thomas (1990) (see data in table~\ref{tab:ix_phase}.)}
\label{fig:ixvel_civ}
\end{figure}

\subsection{V3885 Sgr}
\label{subsec:v3885sgr1}

V3885 Sgr's apparent magnitude is typically $m_{\rm{v}}=10.4$
\cite{93hollander}.  \iue\ low-resolution spectra of V3885 Sgr have been
analysed by Woods et al.~\shortcite{92woods}, who found evidence 
for variation in the N\Vl1240~resonance line in phase
with the C\IVl1549~line. As for IX Vel, Prinja \&
Rosen~\shortcite{95prinja} remarked on evidence for narrow absorption
components in the C\IVl1549~and N\Vl1240~line
profiles.

Cowley, Crampton \& Hesser (1977) derived a 5\fh04 orbital period from
radial velocity variations of Balmer absorption lines. Haug and
Drechsel (1985) used the same technique to derive a zero-phase
ephemeris, $T_0=2,445,148.5535$JD and period, $P=6\fh216$. In their
study this was the most probable period yielded by a power spectrum
analysis, but other periods were found to be only one per cent less
likely. Metz (1989) quotes a period of 5\fh191, but the experimental
technique used to obtain this is not described. In the literature this
last period is most often quoted, but it is not certain why this is
the case. Given the uncertainty in the orbital parameters of V3885 Sgr, 
the spectra of V3885~Sgr are presented without any orbital radial
velocity correction.

\begin{figure*}
\vspace*{10cm}
\includegraphics{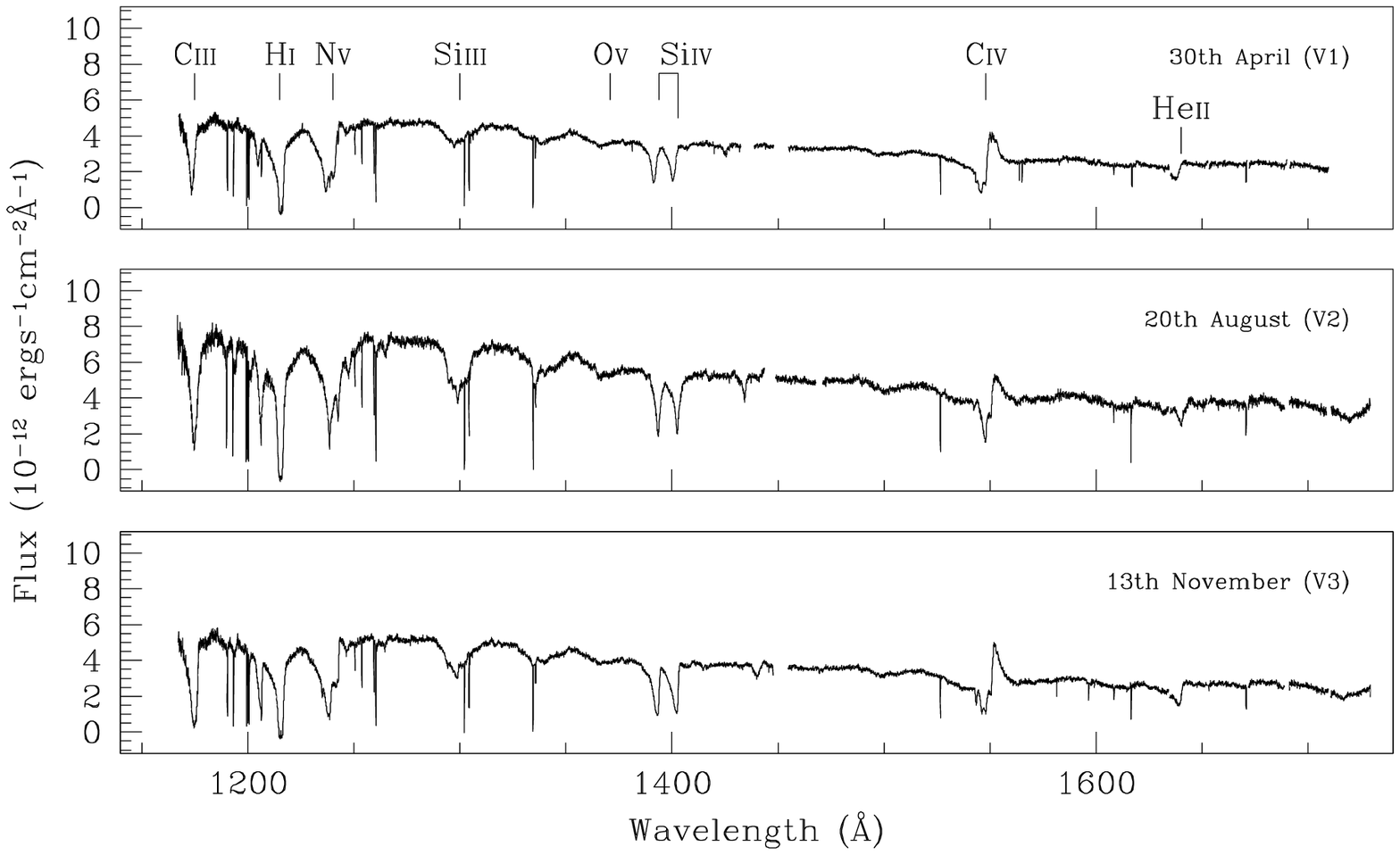}
\caption{The summed spectrum for the (from top to bottom) 30th April,
20th August and 13th November 2000 observations of V3885
Sgr. Resonance transitions and other strong lines are indicated.}
\label{fig:v3885sgr}
\end{figure*}

Figure \ref{fig:v3885sgr} shows the summed spectrum obtained from each
visit to the target (hereafter referred to as V1, V2 and V3). Like IX
Vel, V3885 Sgr's spectrum is characterised by broad blueshifted
absorption and some redshifted emission. Table \ref{tab:v3885_vels}
lists the power law indices and the average flux in the range
1260--1270\thinspace\AA, calculated from a continuum fit to each spectrum, using
the same method as for IX~Vel (section \ref{subsec:ixvel1}). V2 is
noteworthy in that the stronger absorption lines display significantly
reduced asymmetry, with profiles that drop steeply to their minima.

\begin{table*}
\begin{minipage}{12.2cm}
\caption{Measurements of continua and wind lines in V3885 Sgr. `Flux' refers to the mean
continuum flux level in the range 1260--1270\thinspace\AA. `Index'
refers to the index of a power law continuum fit to the data. The
quantities EW and $v^{\rm abs}_{\rm max}$ both depend on eye estimates
of the background continuum level. They are respectively, the
equivalent width of the absorption component and the maximum velocity
of blueshifted absorption. The EW measurement of \lya\ includes the
interstellar component in every case. Where `$s$' appears after an EW
measurement, the measured profile is symmetric, rather than obviously
dominated by blueshifted absorption.  A colon after a velocity
indicates that the measurement is highly uncertain. The typical error
on the equivalent width measurements is 6 per cent.}
\label{tab:v3885_vels}
\begin{tabular}{@{}l|cc|cc|cc}
&\multicolumn{2}{c}{30th April} &\multicolumn{2}{c}{20thAugust}
&\multicolumn{2}{c}{13th November} \\
&\multicolumn{2}{c}{V1 } &\multicolumn{2}{c}{V2}
&\multicolumn{2}{c}{V3} \\
\hline
Flux&\multicolumn{2}{c}{$4.7\pm0.2\times10^{-12}$}&\multicolumn{2}{c}{$7.0\pm0.4\times10^{-12}$}&\multicolumn{2}{c}{$5.0\pm0.2\times10^{-12}$}\\
Index &\multicolumn{2}{c}{$-2.31\pm0.10$}&\multicolumn{2}{c}{$-2.28\pm0.10$}
&\multicolumn{2}{c}{$-2.30\pm0.10$} \\\hline
Lines:		&EW (\AA)	&$v^{\rm{abs}}_{\rm{max}}$	&EW (\AA)	&$v^{\rm{abs}}_{\rm{max}}$	&EW (\AA)	&$v^{\rm{abs}}_{\rm{max}}$	\\
\hline
Si\III$\lambda1206$	&$0.7$		&$-850:$		&$1.2^s$		&$-700:$		&$1.3$		&$-850:$		\\
C\III$\lambda1175$	&$2.3$		&$-2000:$	&$3.2^s$		&$-2000:$	&$3.8$		&$-2000:$	\\
\lya$\lambda1216$	&$7.1$		&$-4000:$	&$6.3^s$		&$-4000:$	&$6.9$		&$-4000:$	\\
N\V$\lambda1240$	&$5.5$		&$-2600\pm200$	&$6.4^s$		&$-2700\pm200$	&$6.4$		&$-2800\pm200$	\\
Si\IV$\lambda1398$&$4.4$		&$-2000\pm200$	&$5.0^s$		&$-1200\pm300$	&$6.3$		&$-2350\pm350$	\\
C\IV$\lambda1549$	&$6.7$		&$-6000:$	&$4.7\:\:$		&$-6000:$	&$6.3$		&$-6000:$	\\
He\II$\lambda1640$&$3.4$		&$-4000:	$	&$2.3$		&$-4000:$	&$3.4$		&$-4000:$	\\
\end{tabular}
\end{minipage}
\end{table*}

The secular changes in the mean spectrum of V3885 Sgr are much like
those seen in IX Vel: we have one dataset exhibiting
well-developed broad blueshifted absorption and narrow superposed dips
(V3, obtained last); a second dataset with somewhat reduced
blueshifted absorption and no superposed dips (V1, obtained first),
and a third set in which the wind signatures weaken noticeably
(V2). We will discuss the observations in this order (V3, V1, V2) so
as to parallel the discussion of IX~Vel presented above. Figure 5
shows a close-up of the major line profiles seen in, respectively, V3,
V1 and V2.

For the first time, in V3885 Sgr's spectrum we see a wind signature in
\lya, although the asymmetry is much weaker than for IX Vel and there
is no redshifted emission component. We also detect blueshifted
absorption in C\IIIl1176, N\Vl1240,
Si\IVl1398~and, unusually, in the Si\IIIl1206 line
that is superimposed on the blue wing of \lya. The mean
C\IVl1549~resonance line profiles are shown in more detail in
figure \ref{fig:v3885_civ}. In V3 and V1, C\IV~blueshifted absorption
extends to at least $-5000\kms$ and redshifted emission reaches to
around 2000\kms. He\IIl1640~absorption extends bluewards to
around $-3000$ to $-4000$\kms, but there is only a barely detectable
emission component.

In V3 the redmost point of absorption in each line profile closely lines 
up with the transition's 
rest wavelength, whereas in V1 this feature is blueshifted. In section 
\ref{subsec:ixvel1} we mentioned that this velocity shift could be
related to the orbital $K$-velocity phase. It is possible that
observation V3 occurred near orbital phase zero or 0.5, while V1
occured when the WD was moving towards us in its orbit.

The continuum flux level is about ten per cent higher in V3 than V1 but, to within
errors, the spectral index is the same (see table \ref{tab:v3885_vels}). 
In V3, the general shape of the line profiles is more asymmetric than in V1. 
In either epoch, the location of the absorption minimum is at about the same 
velocity for all the lines: at around $-250$\kms~for V3 and around $-500$\kms\ 
for V1. This difference could well be due to the, as yet, unknown radial 
velocity shift between the two datasets.

V2 possesses the highest recorded continuum flux level, yet its line
profiles retain the weakest wind signature. C\IIIl1176, \lya, N\Vl1240
and Si\IVl1398 show little, if any, blueshifted absorption. The
C\IVl1549 emission peak is $<50$ per cent (relative to the continuum)
of the peak emission in the other two datasets. However it is also
fair to characterise V3885 Sgr as apparently powering less mass loss
at this time.  The participation of the He\IIl1640 in this weakening
(albeit a less dramatic effect than in IX Vel, I3) suggests that an
upward shift in wind ionization, alone, cannot explain the changed
character of the spectrum.  We shall show in section 4.2 that the
magnitude of the effect seen in V2 is enhanced by some shorter term
variability.  Some lines that are not seen in V1 and are only very
weak in V3 can be clearly seen in V2.  Note, for example, the lines
superimposed on the blue wing of \lya\ between 1187 and 1203 A
and the absorption line at 1543 A.

\begin{figure*}
\vbox to220mm{
\includegraphics{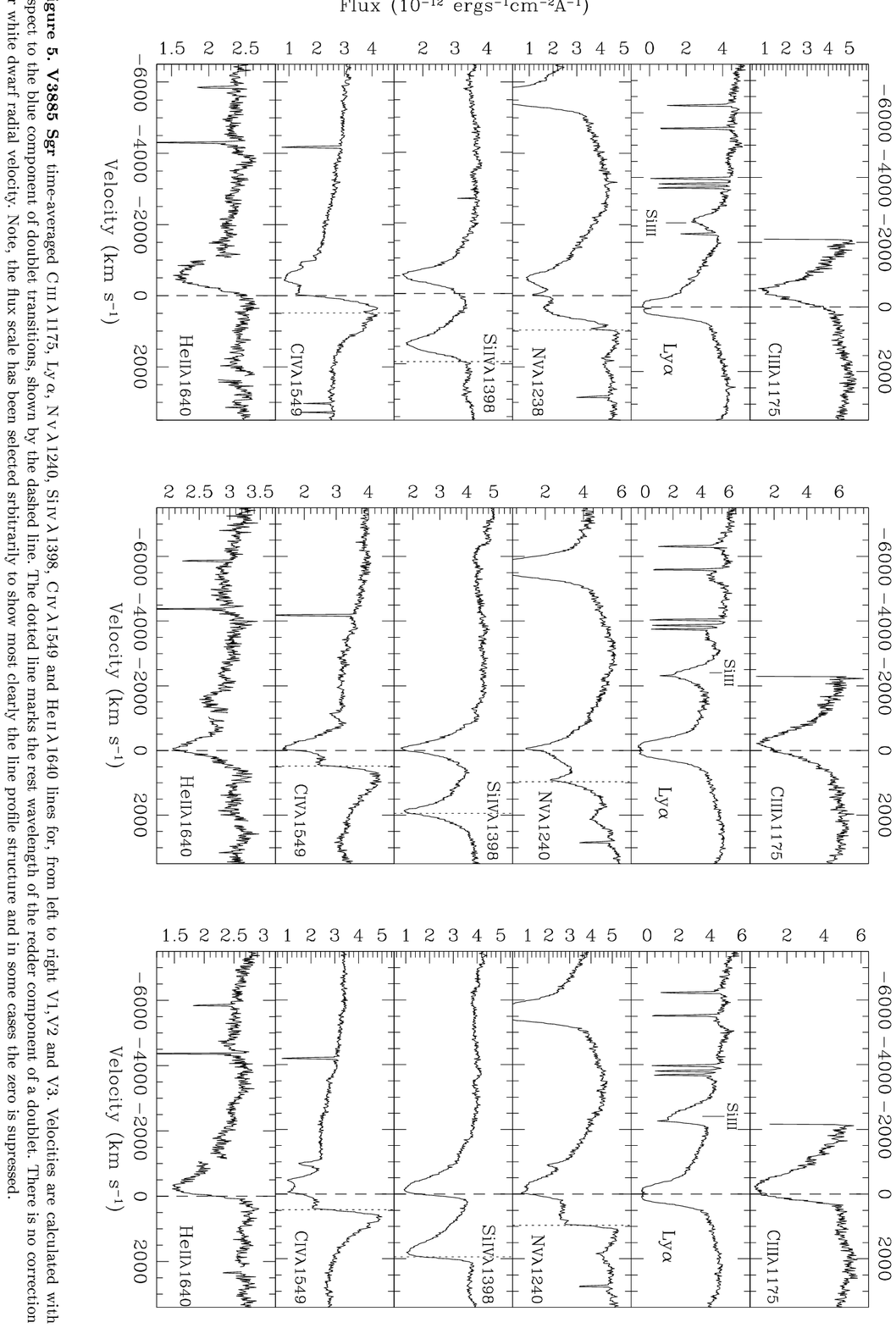}
}
\label{landfig2}
\end{figure*}
\setcounter{figure}{5}

\begin{figure}
\vspace{13cm}
\includegraphics{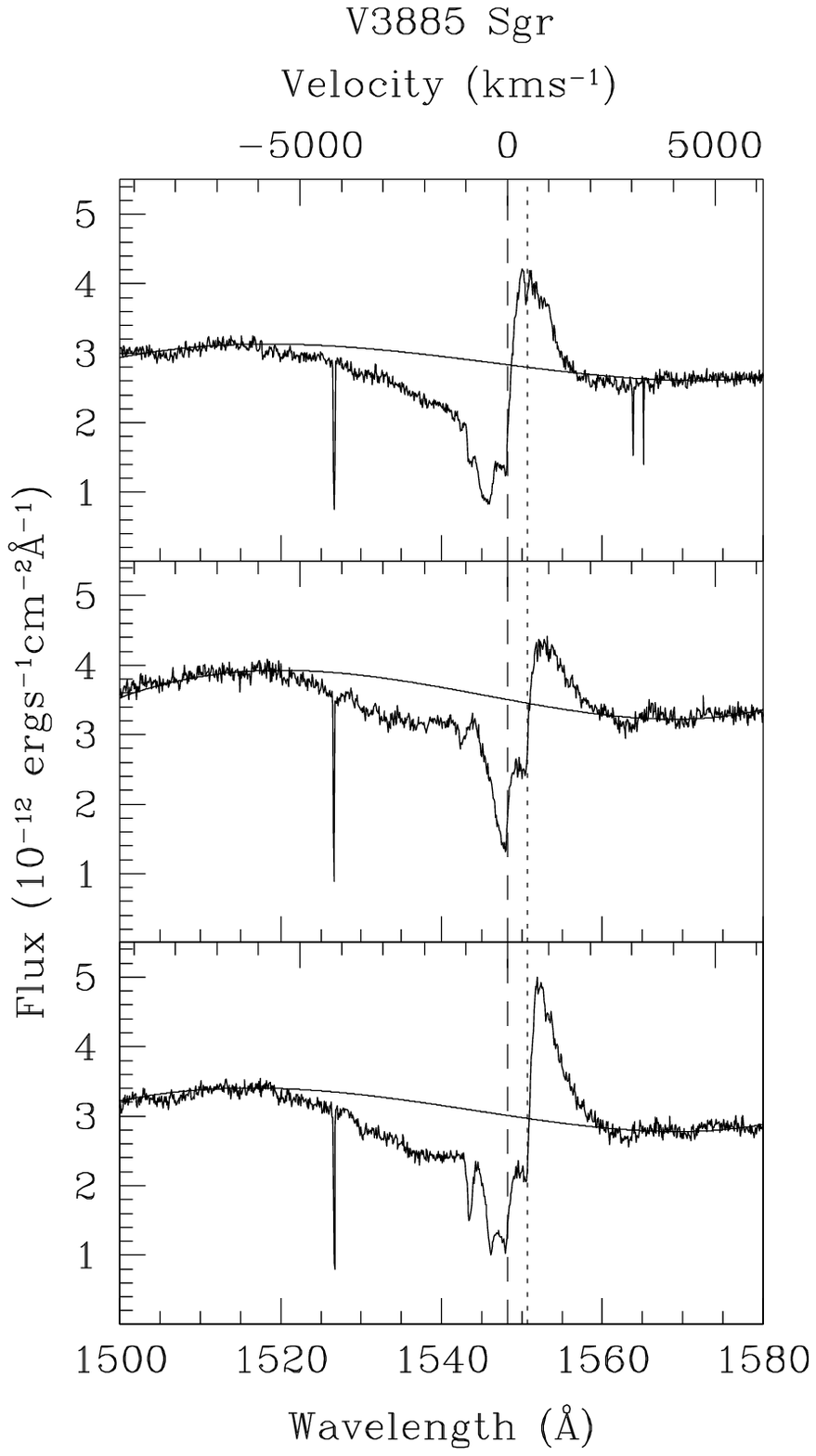}
\caption{Time-averaged C\IVl1549 for, from top to bottom, V1, V2
and V3. Velocities are calculated with respect to the blue component
(1548.2\thinspace\AA) of the doublet, shown by the vertical dashed line. The dotted 
line marks the rest wavelength of the 1550.7-\AA\ component. The narrow 
feature at 1527\thinspace\AA\ is interstellar Si\II. The smooth solid line is a 
4th order fit to the continuum in the vicinity of the C\IV\ line itself 
(included only to give an impression of the full width of the line profile).}
\label{fig:v3885_civ}
\end{figure}


\subsection{The superposed narrow absorption features}
\label{sec:narrow}

The high dispersion of the datasets presents an excellent opportunity for
studying fine structure in wind-shaped line profiles. Both
IX~Vel and V3885~Sgr exhibit very narrow features of $\la100$\kms\
breadth, superposed on the broad blueshifted absorption lines, some of the 
time. These
features are seen clearly in only I1 (IX~Vel) and V3 (V3885~Sgr) -- the two
datasets that also display the highest equivalent width of blueshifted
absorption.  They also appear fleetingly in V1 (see below and
section \ref{sec:timevar}).  The absence of these narrow features in
I2, I3 or V2, hints at a positive correlation between their appearance and
the overall wind column as measured by total UV blueshifted absorption
equivalent width.

\begin{table}
\caption{The velocity offsets of narrow absorption features superposed on broad
blueshifted absorption profiles.  All velocities are given in \kms~with
respect to the rest wavelength of the transition in the heliocentric
frame}
\label{tab:nabs_vels}
\begin{tabular}{@{}lllll}
\hline
		&\multicolumn{2}{c}{IX Vel (I1)}	&\multicolumn{2}{c}{V3885 Sgr (V3)}	\\
Transition & \multicolumn{2}{c}{Velocity (km s$^{-1}$)} & \multicolumn{2}{c}{Velocity (km s$^{-1}$)} \\
\hline
\lya			&$-860\pm60$	&		&		&\\
N\Vl1238.8	&$-890\pm80$	&		&$-900\pm70$	&$-60\pm50$	\\
N\Vl1242.8	&		&		&		&$-70\pm50$	\\
Si\IVl1393.8	&$-850\pm50$	&$-20\pm60$	&$-850\pm50$	&\\
Si\IVl1402.8	&		&$\,\,\,\,\,\,0\pm40$&		\\
C\IVl1548.2	&$-890\pm100$	&		&$-910\pm120$	&$-50\pm40$	\\
C\IVl1550.7	&$-890\pm70$	&		&$-900\pm60$	&$-70\pm50$
\end{tabular}
\end{table}

Measured velocities for the narrow absorption components seen in IX~Vel (I1)
and V3885~Sgr (V3) are given in table~\ref{tab:nabs_vels}.  For consistency
the velocities given for IX~Vel are not corrected for putative WD
motion and given as observed, as they are for V3885~Sgr.  We therefore
remind that the correction to the WD rest frame for I1 implied by
data from Beuermann \& Thomas (1990) is to blueshift the quoted velocities
by a further 173$\pm$5~\kms.  We suspect that the correction for V3 onto the
WD rest frame is small given the near alignment in the
N\Vl1240 and C\IVl1549 profiles of the red absorption edges
with the doublet rest wavelengths in the heliocentric frame (see also section
\ref{subsec:v3885sgr1}).

In IX Vel, I1, it is striking that the same velocity of roughly $-900$\kms
~is picked out by narrow absorption features in four transitions, including
\lya~at one ionization extreme and N\Vl1240 at the other.
Coincidentally, in the spectrum of V3885 Sgr, V3, the blueshifted narrow
absorption dips again cluster around a velocity of about $-$900~\kms.
It is true of both I1 and V3,
that the dips are most prominent in the resonant C\IV~and N\V~transitions and
essentially absent from the subordinate C\IIIl1175 line.

The only other narrow velocity component clearly seen in both I1 and V3
is, in both cases, located close to line centre.  In I1, such a feature is
apparent in both doublet components of Si\IVl1398 at a blueshift of
just 20~\kms.  Because this feature is rather too wide to be interstellar
(with a FWHM of $\sim$60 \kms) and because it shows some fading during the
course of the observation, it must be identified as intrinsic to IX~Vel.
In V3 the
affected transitions are the N\Vl1240 and C\IVl1549 doublets,
showing dips at an offset of about $-60$~\kms , again with a FWHM of 
$\sim$60~\kms .
By the same reasoning these are most likely intrinsic to V3885~Sgr.

V1, the first of the V3885~Sgr observations, also betrays traces of narrow
absorption features at about $-900$\kms .  Specifically these are located
in \lya~at $-890\pm50$\kms~and Si\IVl1398, at $-870\pm30$\kms. 
C\IVl1549 has a trace
of two narrow dips at $-880\pm50$\kms, relative to each doublet component,
and also a dip at $-1140\pm30$\kms, relative only to the bluer component.
These features are only present for the first 600--700\thinspace sec of the
observation.

\section{Variability on timescales shorter than $\sim$2000 sec.}
\label{sec:timevar}

An analysis of time-variability of the resonance
lines during each observation is presented in this section. 
The N\Vl1240, Si\IVl1398~and
C\IVl1549~resonance lines are displayed as trailed mean-subtracted 
spectra, and, where their inclusion clarifies the discussion, as trailed 
normalised spectra, as well.

To reduce the prominence of noise in the trailed spectra, the 1D
spectra were smoothed by convolution with a gaussian function of either
$\sigma=0.44$\thinspace\AA\ (4 pixels, I1 and V3) or $\sigma=1.1$\thinspace\AA\
(10 pixels, V1 and V2).
The spectra were then normalized by fitting a 4th order chebyshev function to
the time-averaged data; the fit was checked to assure that it did not
produce strong curvature in the vicinity of the spectral lines.
Examples of the local behaviour of these fits are shown in figures 
\ref{fig:ixvel_civ} and \ref{fig:v3885_civ}. All time-resolved 
spectra were divided by this continuum fit and then rescaled to ensure
a continuum level of 1. The difference spectra were
prepared for plotting by subtracting the normalized time-averaged spectrum
from each time-bin's normalized spectrum.


\subsection{IX Vel}
\label{subsec:ixvel_2}

Only the I1 data, which display the strongest wind profiles and highest flux 
level, show any sign of time-variability in the trailed mean-subtracted 
spectrum and the normalised spectrum, so we present trailed spectra only for 
this \hst\ visit. IX Vel's UV brightness brings it close to the 
brightness limit of the detectors: this meant that at intervals during the 
observation the data buffer reached its limit and stopped recording data. 
This results in the loss of approximately 10\thinspace sec of data every 
100\thinspace
sec, after an initial 570\thinspace sec with 
no gaps. However, normalisation of the data ensures that there is no 
continuum variation resulting from this. We concern ourselves with time 
bins of 30\thinspace sec.

\begin{figure*}
\vspace*{9cm}
\includegraphics{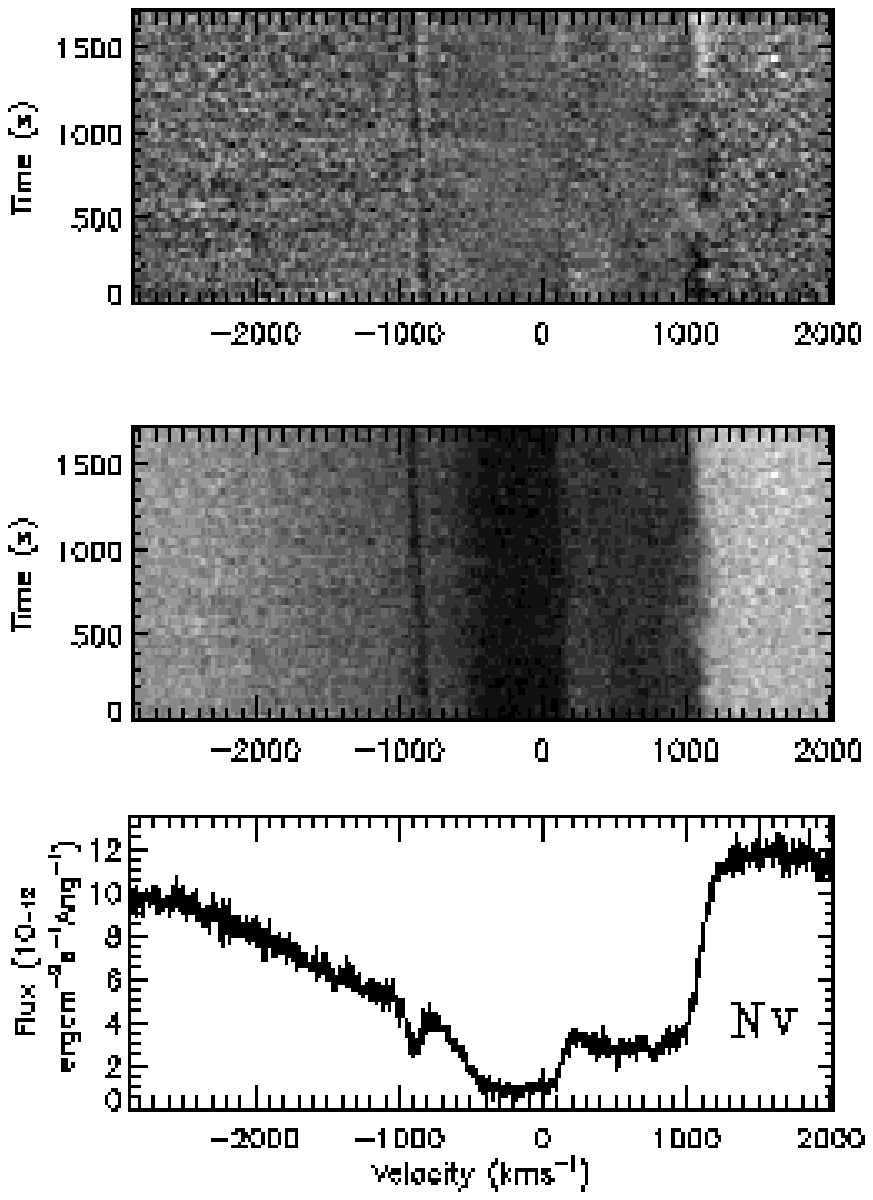}
\includegraphics{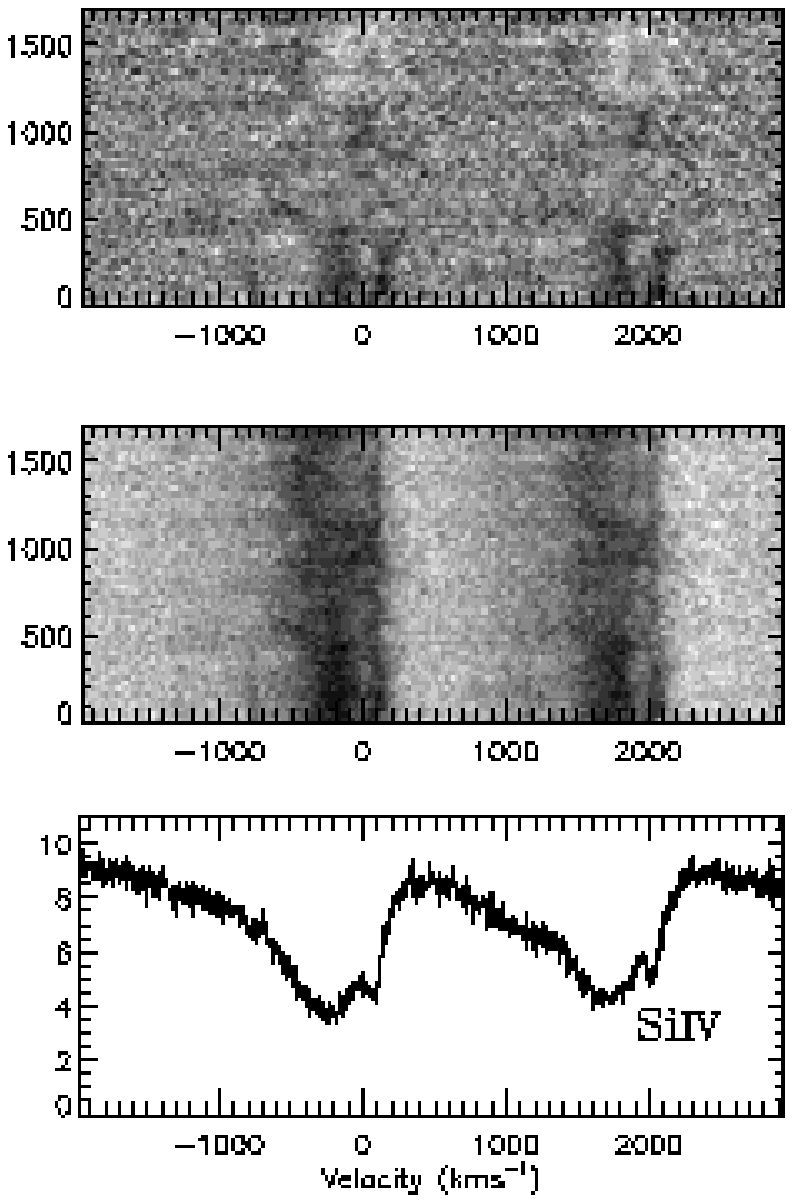}
\includegraphics{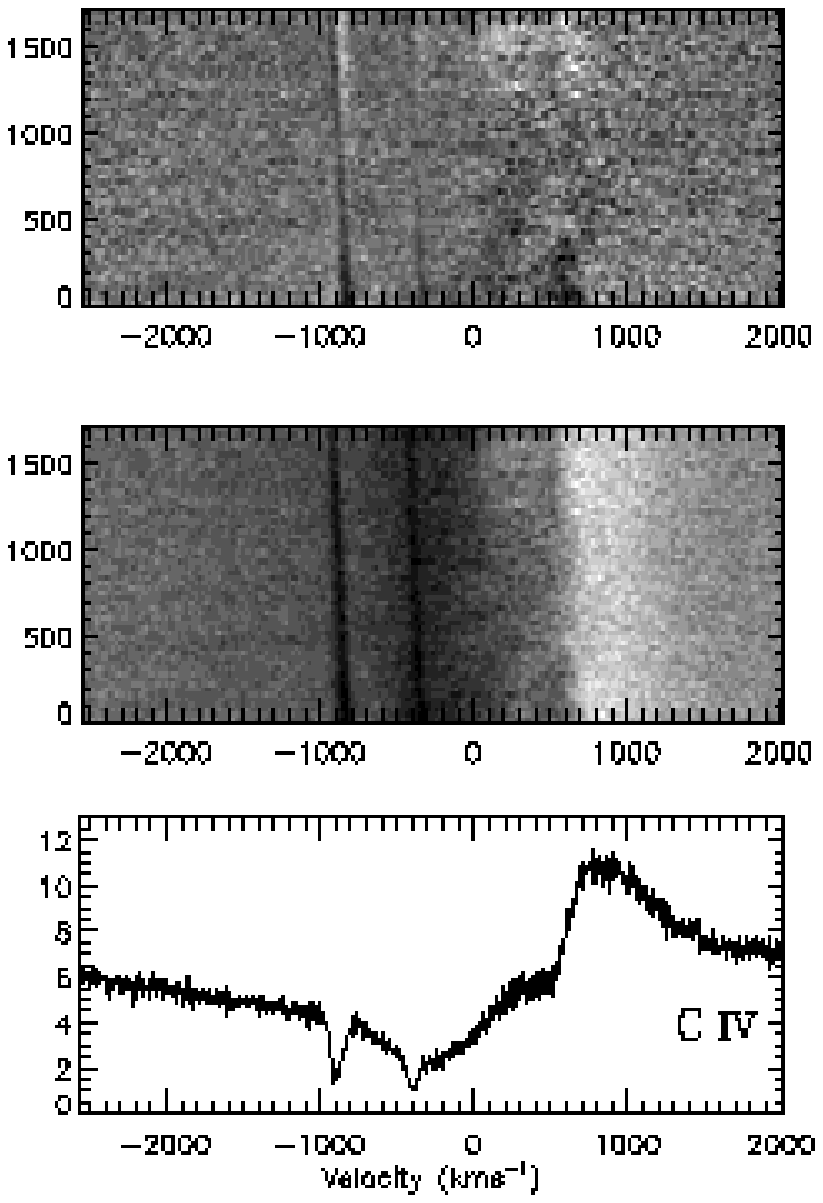}
\caption{Trailed mean-subtracted spectra (top), trailed 
normalised spectra (middle) and time-averaged spectrum (bottom) of, from left 
to right, the N\Vl1240, Si\IVl1398~and C\IVl1549~for 
IX Vel, I1. The spectrum was smoothed with a gaussian function of
$\sigma=0.44\,$\AA. The greyscale range for the mean-subtracted spectra is as follows: N\V, -0.19 to 
0.19; Si\IV, -0.15 to 0.18; C\IV, -0.24 to 0.27. Darker grey indicates a flux 
of less than average, lighter grey to white indicates a flux greater than 
average.}\label{fig:ixgr1}
\end{figure*}

In figure \ref{fig:ixgr1} the narrow absorption features noted in
section \ref{subsec:ixvel1} are apparent in the difference spectrum
as narrow dark features moving bluewards. In the
N\Vl1240~profile only the narrow feature that corresponds to
the 1238.8-\AA\ component of the doublet can be seen. Its position
shifts from $-840$ to $-930\;(\pm20)\;$\kms. The redder component
falls at approximately the rest wavelength of the bluer feature, so
its presence is obscured by variability noted there (see below). Two
narrow features at the doublet separation move along the 
C\IVl1549~profile from $-850$
to $-930\;(\pm20)\;$\kms~and from $-370$ to $-450\;(\pm20)\;$\kms,
consistent with the change in the N\V~narrow feature. This is slightly
more than the change in velocity expected from the drift with orbital
phase (see table \ref{tab:ix_phase}), which we have estimated at
$\sim60$\kms. In the case of the bluer C\IV~narrow feature, the
absorption strength decreases by about 50 per cent from beginning to end. The
other features are more obscured by noise, but they can also be seen
to decrease in strength during the observation.

In both N\V~and C\IV, variability is seen near line centre. 
In the normalised trailed spectra
this is seen as blueshifting of the redmost point of absorption. This
shift is measured, through inspection of the 1D spectrum from each 30-sec
exposure, to be 90\kms. This is of the same magnitude as the movement
of the narrow features. The correlation between the velocity shift of
these two independent features (narrow absorption lines at
$\sim-900\kms$ and the redmost point of absorption) suggests that the
predicted orbital phase is about right and that these
two features are stationary in the WD rest frame. By contrast, in I2
and I3 there is no shift in the position of the redmost absorption
point during the observation.  From the predicted orbital
phases (table \ref{tab:ix_phase}) we would expect to see less of a
velocity shift than in I1.

In approximately the first 250{\thinspace}s of I1, a narrow absorption
feature is visible superimposed on the Si\IVl1398 doublet at
$\sim-850\pm25$\kms. In figure \ref{fig:ixgr1} these features are just
discernable as a narrow darker feature moving bluewards. There are
other narrow absorption features near rest that
are only clearly visible for the first 600\thinspace
sec or so, after which, they appear intermittently. These serve to confuse
the blueshifting redmost absorption limit that also seems to be 
present in Si\IV.

No velocity shift has been detected in the disk-formed absorption lines from
1300 to 1400\thinspace\AA\ (Si\IIIl1300/O\Il1304,
C\II$\lambda1335$ and O\V$\lambda1371$ are the dominant species). However, 
these lines are broad and shallow with many components and so, unlike the 
wind-formed lines, tend to thwart efforts to identify shifts by means
of cross-correlation -- whether these lines shift with the WD
radial velocity remains moot.

\subsection{V3885 Sgr}
\label{subsec:v3885sgr_2}

\begin{figure*}
\vspace*{9cm}
\includegraphics{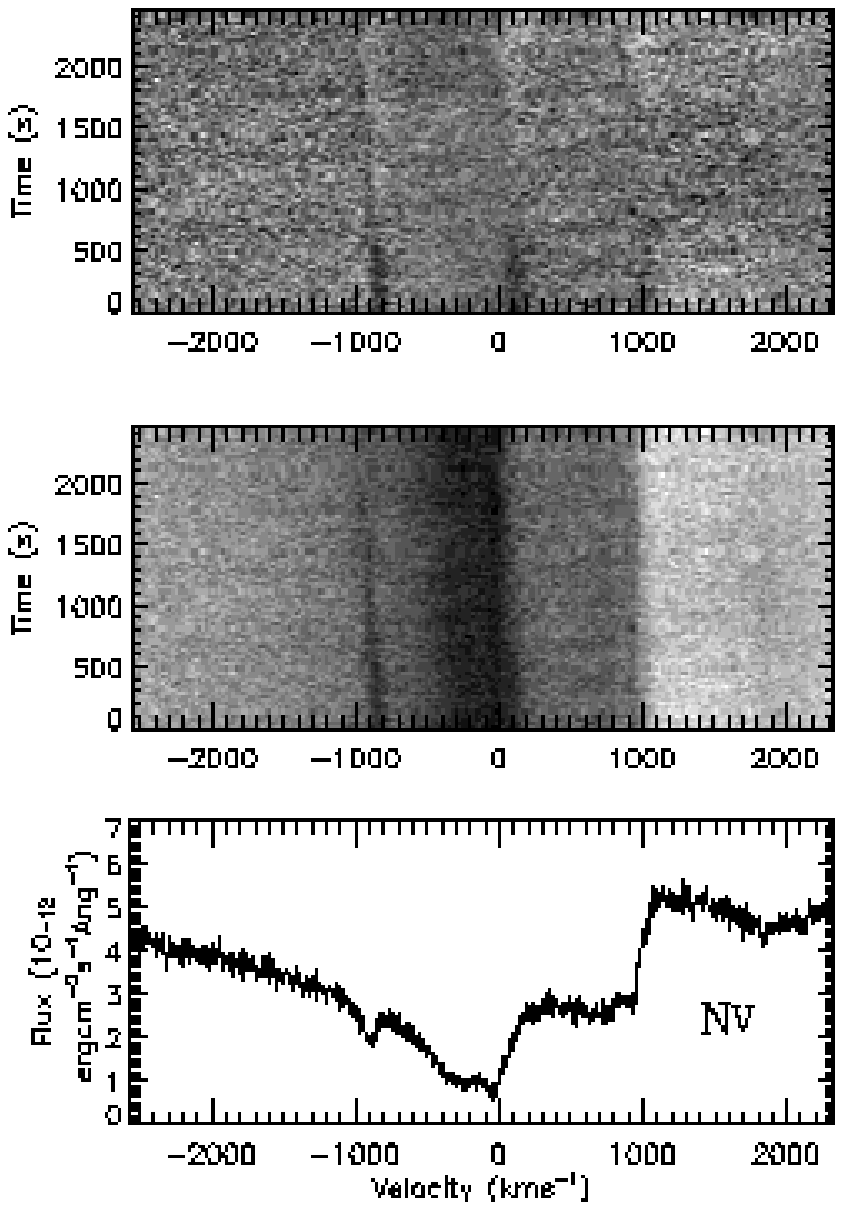}
\includegraphics{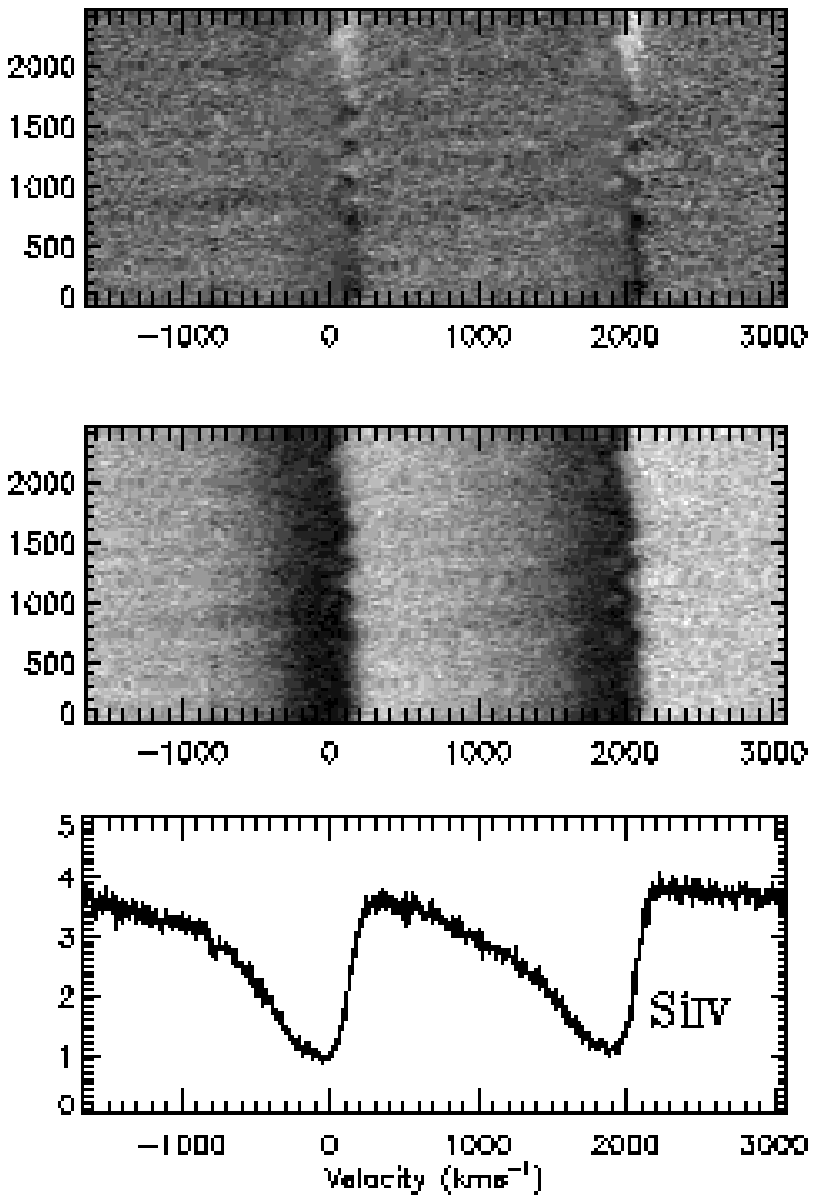}
\includegraphics{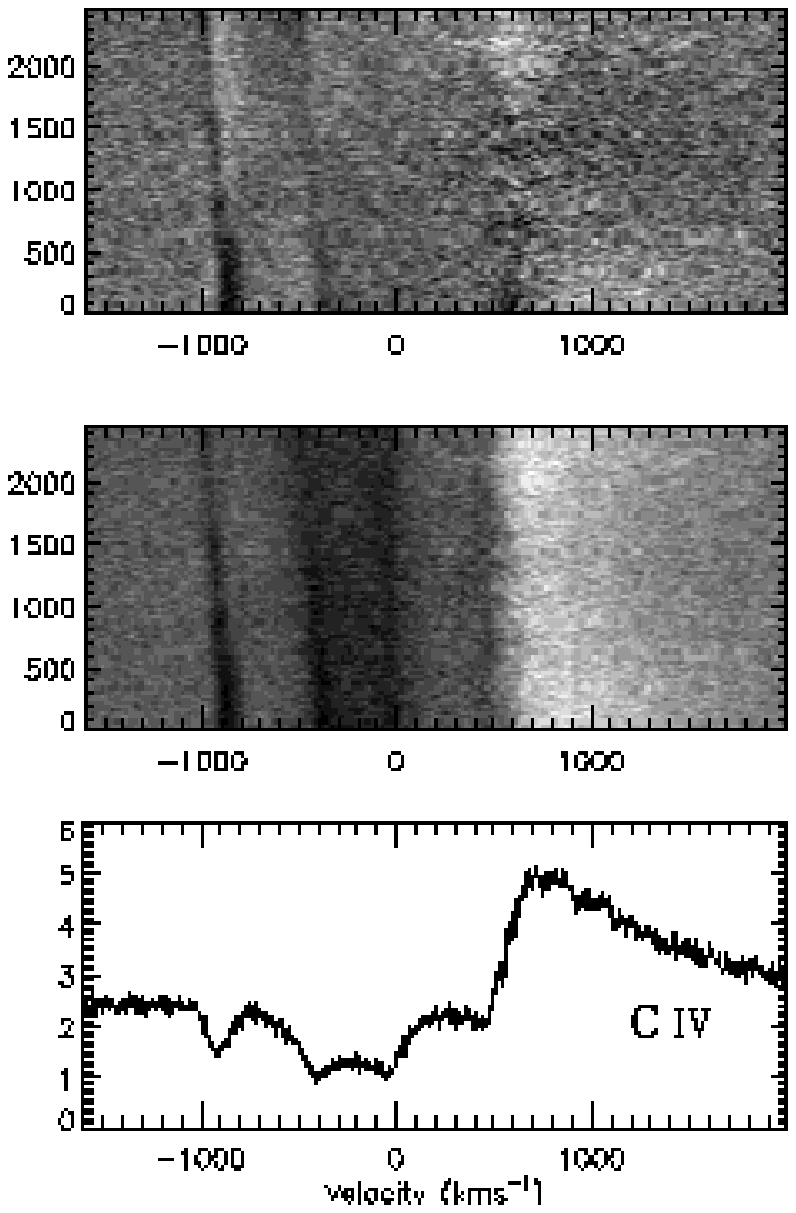}
\caption{Trailed mean-subtracted spectra (top panel), 
trailed normalised spectra (middle) and time-averaged spectrum
(bottom) of, from left to right, the N\Vl1240,
Si\IVl1398~and C\IVl1549 lines for V3885 Sgr, V3. The spectrum was smoothed with a gaussian function of
$\sigma=0.44\,$\AA. The greyscale range for
the mean-subtracted spectra is as follows: N\V, -0.17 to 0.23; Si\IV,
-0.12 to 0.17; C\IV, -0.34 to 0.42.  Darker grey indicates a flux of
less than average, lighter grey to white indicates a flux greater than
average.}
\label{fig:v3885gr3}
\end{figure*}

\begin{figure}
\vspace{14.5cm}
\includegraphics{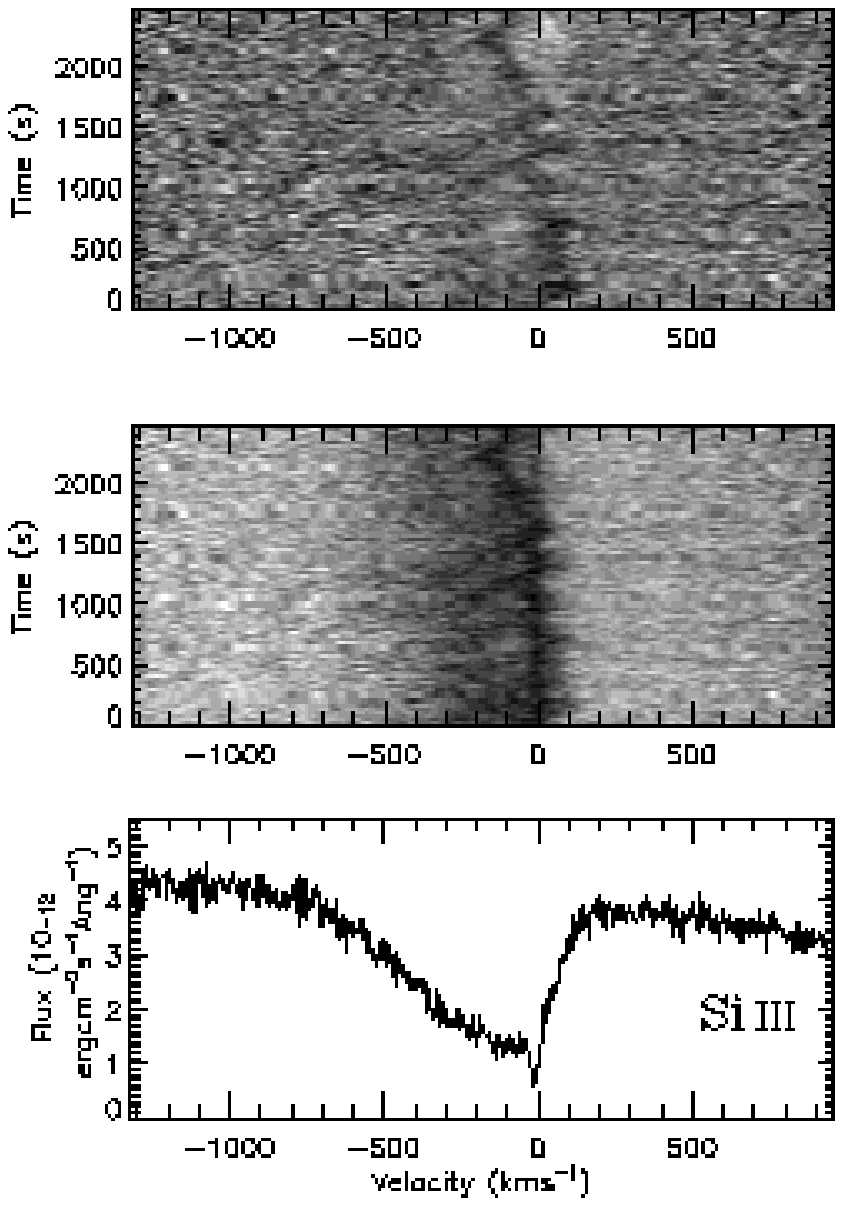}
\caption{Trailed mean-subtracted spectra (top panel), trailed
normalised spectra (middle) and time-averaged spectrum (bottom) of the
Si\IIIl1206 line for V3885 Sgr, V3. The spectrum was smoothed with a gaussian function of
$\sigma=0.44\,$\AA. The greyscale range for the mean-subtracted
difference spectra is -0.19 to 0.25. Darker grey indicates a flux of
less than average, lighter grey to white indicates a flux greater than
average.}
\label{fig:v3grsiiii}
\end{figure}

Presented in figures \ref{fig:v3885gr3}, and \ref{fig:v3grsiiii} are
the V3 time series for the N\Vl1240, Si\IVl1398,
C\IVl1549~lines and Si\IIIl1206 transitions.  This was
the epoch of greatest wind activity in which narrow blueshifted
absorption features, like those in IX Vel's I1 spectrum (figure
\ref{fig:ixgr1}), were seen superposed on the broad absoption profile. We begin here.

In figure \ref{fig:v3885gr3} a narrow absorption feature corresponding
to the 1238.8-\AA\ component of the N\V~doublet moves bluewards from
approximately $-860$\kms\ to approximately $-1100$\kms , becoming less
prominent towards the end of the observation. This behaviour is
repeated in the C\IV~line, which displays two blueshifting narrow
absorption features, moving from $-890$ to $-1020\;(\pm20)$\kms~and
$-380$ to $-520\;(\pm40)$\kms. These narrow features become weaker
during the course of the observation, decreasing in strength by
approximately 40 per cent. The narrow line features noted just shortward of
the N\V~and C\IV~doublet rest wavelengths in section
\ref{subsec:v3885sgr1} do not appear in the mean-subtracted spectra,
indicating that they do not vary.  Also in common with IX~Vel, I1, the
normalised greyscales of the V3 time series show blueshifting of the
redward limit of the absorption. We measure this shift to be between
65 and 80~\kms and note it is present in all lines shown. Its
magnitude is rather less than the $\sim$140 \kms\ shift apparent from
the superposed narrow absorption lines.

If we assume that the movement of the narrow absorption features is
linked to binary orbital motion, as was suggested for IX Vel (see
section \ref{subsec:ixvel_2}), we can use the increasing blueshift
observed to put V3885 Sgr in either the first or fourth quarter of its
orbit.  Haug \& Drechsel (1985) derive a $K$-velocity of 274\kms\ from
Balmer emission lines. This would give a minimum velocity shift during
the period of the observation of between 90\kms~and 203\kms, a range
that includes the velocity shift measured.

In the Si\IV\ and C\IV~line profiles, we note a $\sim200$~\kms\ fluctuation in
the position of the red absorption edge: it
wobbles, moving bluewards then redwards on a timescale of about
350\thinspace sec. The same fluctuation is even more clearly present in the
Si\IIIl1206 line, shown in figure \ref{fig:v3grsiiii}. The fluctuation occurs in
phase and with approximately the same velocity amplitude in all three
transitions.  It is strongest around the middle of the V3 observation
and decays away noticeably towards the end.  Time-varying structure,
other than the line centre wobble, cannot be seen in the C\IV\
profile. A related variation may be superposed on the
blueshifted Si\IV\ absorption, and there is a definite indication of some
streaking in the weaker, less opaque Si\III\ line greyscales.
Assuming that the line-centre wobble and the blueshifted streaking are
aspects of the same phenomenon, we may deduce that there is some fine
structure present in the outflow.  The absence of any variability
redward of line centre makes it unlikely that the rotating disk (or
WD) photosphere is implicated.  If 350\thinspace sec is comparable with
the local dynamical time, this timescale of variation implies a
characteristic disk radius of about 10 WD radii (for
a 1\msun WD).

\begin{figure*}
\vspace*{6cm}
\includegraphics{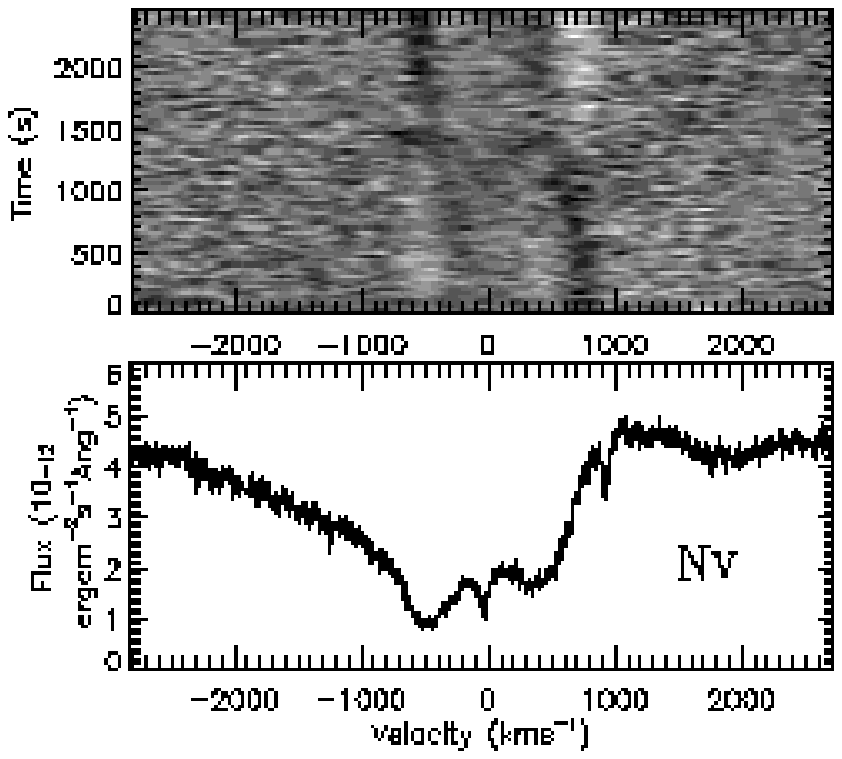}
\includegraphics{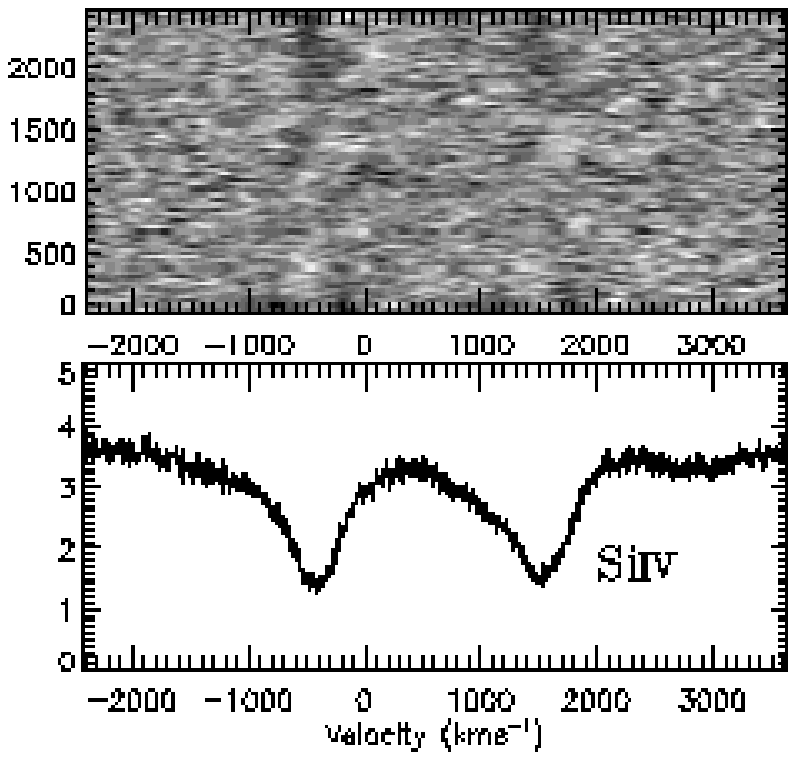}
\includegraphics{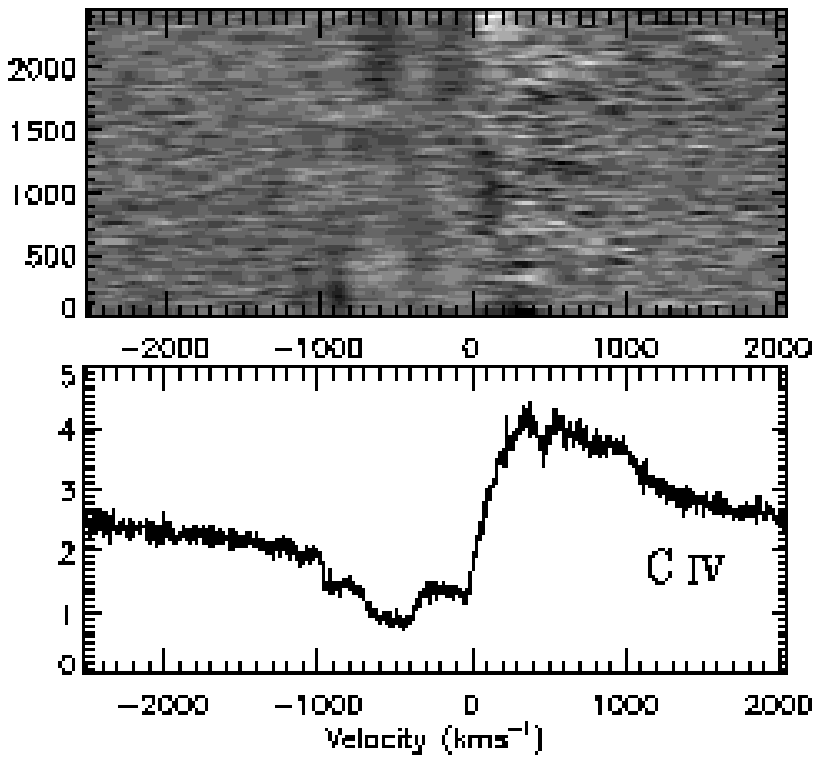}
\caption{Greyscale difference spectrum of, from left to 
right, the N\Vl1240, Si\IVl1398~and C\IVl1549~line for V3885 Sgr, V1,
shown with the time-averaged 1D spectrum. The spectrum was smoothed with a gaussian function of
$\sigma=1.1\,$\AA. The greyscale range for the mean subtracted
difference spectra is: N\V, -0.16 to 0.13; Si\IV, -0.10 to 
0.14; C\IV, -0.16 to 0.27. Darker grey indicates a flux of less than average, 
lighter grey to white indicates a flux greater than average.}
\label{fig:v3885gr1}
\end{figure*}

Figure \ref{fig:v3885gr1} shows the greyscale line profiles for the
first V3885~Sgr observation, V1. Two absorption features,
approximately 200\kms~wide, move bluewards parallel to each other,
shifting by approximately $-300$\kms~during the observation. This is
most evident in the N\Vl1240 line profile.  These changes are
not linked to the narrow features at approximately the rest
wavelengths of the doublet, as they are stationary throughout the
observation.  Instead, they are a sign of a shifting (but always
blueshifted) absorption minimum against a background of an otherwise
invariant broader profile.  The net effect is that the line profiles
show a slightly stronger blue skew towards the end of the observing
window.  This pattern of behaviour is distinct from the blueshifting
of the redmost limit of absorption, as described for V3, wherein the
line profile shape as a whole remains constant.

Towards the beginning of the observation the C\IV~line displays narrow
absorption minima at $-1140\pm30$\kms~, which persists for about
200\thinspace sec,
and at $-900\pm50$\kms, which disappears after about 700\thinspace sec. The
features are harder to discern than the narrow features seen in V3 as
they are broader and shallower with respect to the overall line
profile. Also present are two blueshifting broader features that
resemble those seen in the N\V~line.

\begin{figure*}
\vspace*{6cm}
\includegraphics{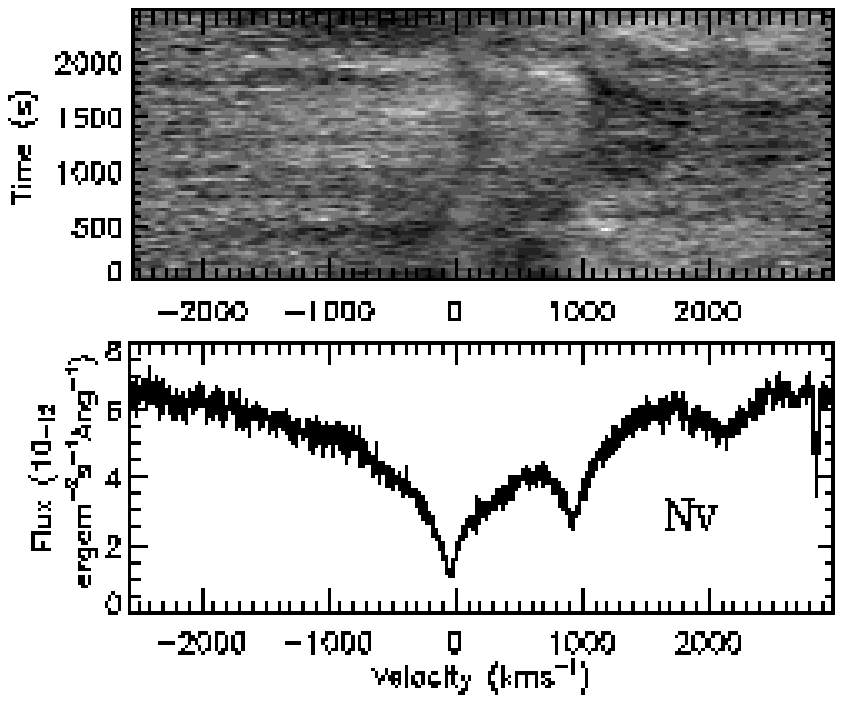}
\includegraphics{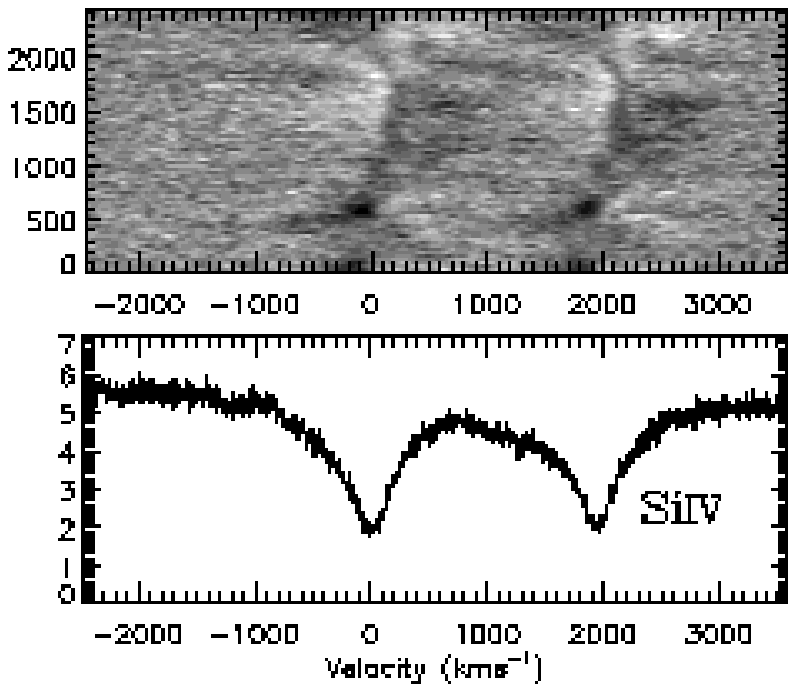}
\includegraphics{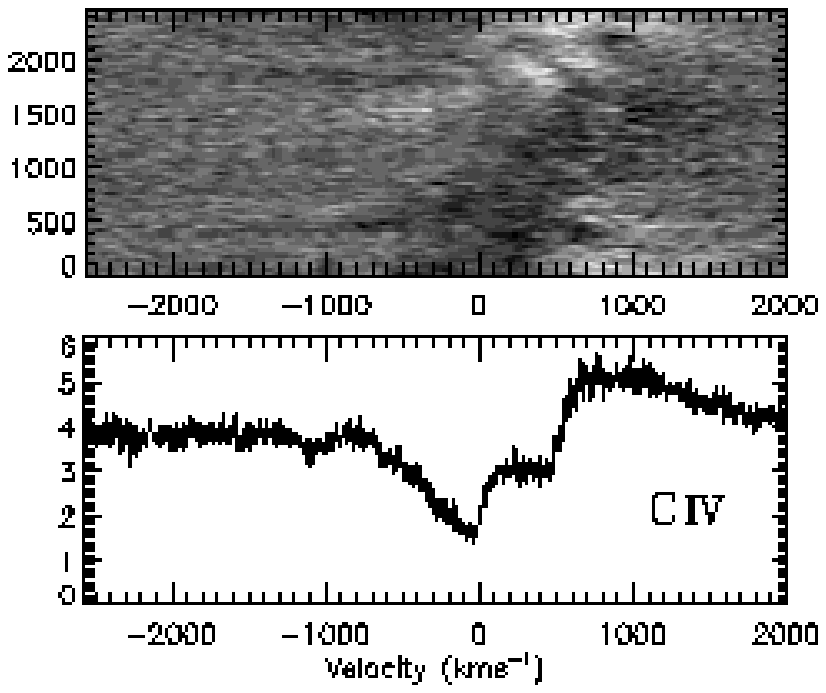}
\caption{Greyscale difference spectrum of, from left to 
right, the N\Vl1240, Si\IVl1398~and C\IVl1549~line 
for V3885 Sgr, V2, shown with the time-averaged 1D spectrum. The spectrum was smoothed with a gaussian function of
$\sigma=1.1\,$\AA. The greyscale range for the mean-subtracted
difference spectra is: N\V, -0.22 to 0.28; 
Si\IV,  -0.24 to 0.23; C\IV, -0.28 to 0.45. Darker grey indicates a flux of 
less than average, lighter grey to white indicates a flux greater than 
average.}
\label{fig:v3885gr2}
\end{figure*}

V2 displays the least distinct wind signatures of the three
observations.  Nevertheless C\IVl1549 retains a clear P-Cygni
character in the mean spectrum that bears a family resemblance to
those for V1 and V3 (e.g. figure~\ref{fig:v3885_civ}).  In this case,
the trailed spectrum reveals a more complex story: in figure
\ref{fig:v3885gr2} we see parabolic curved absorption features that
reach a maximum velocity at around half way through the
observation. This behaviour is most evident in the N\V~and Si\IV~
trailed spectra -- probably because there is rather less masking broad
blueshifted absorption. The movement of the absorption shows too much
curvature to be explained by orbital motion. Instead, it can be viewed
as a result of a change from a windy, asymmetric absorption profile to
a more centred, symmetric profile and back again.

\begin{figure*}
\vspace*{13.5cm}
\includegraphics{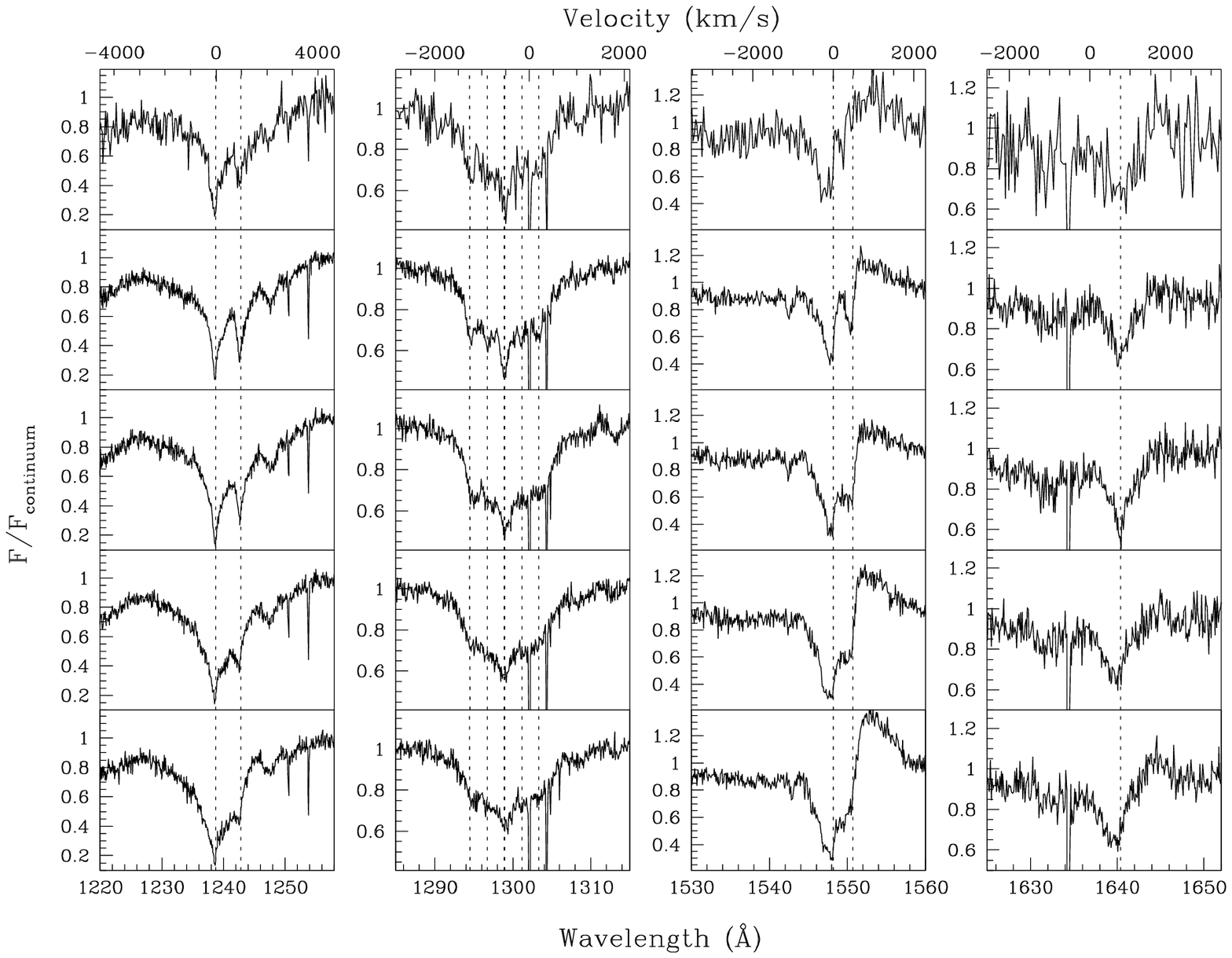}
\caption{The profiles of the (from left to right) 
N\V$\lambda1240$, Si\III$\lambda1300$, C\IV$\lambda1549$~and 
He\II$\lambda1640$~lines in $500$-sec time-bins (from bottom to
top) for V3885 Sgr, V2. The vertical dotted lines mark the transition rest wavelengths.}
\label{fig:v3885_lines}
\end{figure*}

To give a clearer impression of this, we have broken the V2
observation down into 500\thinspace sec time slices and plotted
selected line profiles from them as figure \ref{fig:v3885_lines}.
These show that V2 begins with blue-skewed absorption profiles in the
N\V\ and He\II\ lines. There is certainly a wind blowing, but
apparently it is not so well-developed as in V1 and V3. Both the NV
and HeII lines then lose what little blueshifted absorption they had
to start with, becoming even more symmetrical 1000 and 2000 seconds
after the start of the observation. At the same time the C\IVl1549
profile shows some sharpening of the blueshifted absorption and some
loss of redshifted emission.  As these changes occurs, the multiplet
components of the presumably disk-formed Si\IIIl1300 line stand out
sharply from the underlying absorption trough.  A question to consider
below is whether this variation during the V2 observation is evidence
of a fading wind or of a change in the visibility of the inner disk
from whence the wind flows.

\section{Discussion}
\label{sec:discus}

In this paper we have presented high-time and wavelength resolution
UV observations of the nova-like variables IX Vel and V3885 Sgr.
We begin the discussion of the outcome of these observations by
reviewing some of the binary parameters and other specifics of the two
target systems (section~\ref{subsec:discus1}).  Next, we evaluate what has
been learned from the time-series observations and compare with the
already published studies of BZ~Cam and V603~Aql 
(section~\ref{subsec:discus2}).  We then reconsider the suitability of the 
line-driven disk wind model in the light of the new and, indeed, some 
older observations (section~\ref{subsec:discus3}).  Section 5.4
contains some comment on the way forward from here.

\subsection{IX~Vel and V3885~Sgr}
\label{subsec:discus1}

First we compare and contrast IX~Vel and V3885~Sgr, as known from the
literature, and as we have found them here.  Sufficiently precise parallaxes
exist for both binaries that we can comment on their likely relative
luminosities.  IX~Vel is listed in the HIPPARCOS catalogue as having
a parallax of 10.38 $\pm$ 0.98 mas, placing it at a distance of
96$^{+11}_{-8}$ pc.  V3885~Sgr has a rather less certain parallax of
9.11$\pm$1.95 mas, suggesting a distance of 110$^{+30}_{-20}$ pc.
Combining these data with the impression from photometry that V3885~Sgr
is typically a magnitude (i.e. close to a factor of 2) fainter at optical
wavelengths, one may deduce that V3885~Sgr is anywhere between about as
optically luminous as IX~Vel and half as luminous.  The difference in
apparent brightness is much the same at UV wavelengths, as measured in
these data (cf. I1 and V3) and indeed in the older \iue\ and more recent \fuse\
data available from archive.  

The orbital inclination of IX~Vel would appear to be relatively settled
at $i=60\pm5\degr$ (Beuermann \& Thomas 1990).  The situation regarding
V3885~Sgr is less clear: all that can be said is that the
binary is not eclipsing and that there is a difference of opinion
in the literature, with Cowley et al. (1977) favouring
$i\la50\degr$ and Haug \& Drechsel (1995) preferring either
$60\degr$ or $70\degr$.  In the absence of better observational
constraints, it would appear that to first order IX~Vel and V3885~Sgr
are at similar inclinations.  A problem for the analysis of our V3885~Sgr
observations has been the absence of a compelling measurement of the
semi-amplitude of the white-dwarf $K$-velocity, and great uncertainty in the
orbital ephemeris.  It is to be hoped that this can be corrected soon by
means of ground-based optical spectroscopy.

Against this background, it is not too surprising that the \hst/STIS UV
spectral characteristics of the two systems are not very different.  In the
mean of the index, the slope of the UV continuum in the two objects is about
the same, fitting the power law $F_{\lambda} \propto \lambda^{-2.3}$, but
we do see more variation in continuum slope in IX~Vel.  Tentatively we find 
that a redder UV continuum in IX~Vel (power law index $-2.3$) is linked 
to more extreme spectral evidence of mass loss -- when the wind is all but
absent a bluer slope of index $-2.7$ is observed.  This leads us to surmise
the presence of some diffuse Balmer continuum emission due to the wind that
acts to redden the spectrum slightly when the wind blows.  If the same is
happening in V3885~Sgr, it may be less evident because the variation in wind
activity is less marked and perhaps (more speculatively) because the binary's
orbit is less highly-inclined to the line of sight.

Both IX Vel and V3885~Sgr show evidence of mass loss in \lya~as well as
in the often-remarked C\IVl1549, Si\IVl1398 and
N\Vl1240 resonant transitions.  V3885~Sgr distinguishes itself from
IX~Vel in also presenting with blueshifted absorption in
Si\IIIl1206 -- this is particularly evident in V3, the third epoch
data wherein the wind spectral signatures were most pronounced.

In general, IX Vel appears to possess the more opaque outflow.
This can be seen by comparing the strength of absorption in the two datasets
that display the most well-developed wind profiles: I1 and V3. In particular,
the two most opaque transitions, N\Vl1240 and C\IVl1549 have
greater equivalent widths in I1 than in V3 (see tables \ref{tab:ixvel_vels}
and \ref{tab:v3885_vels}).  Despite this evidence of a difference in typical
wind optical depths, it cannot yet be determined whether this is the effect
of a greater global mass loss rate or simply an orbital inclination effect. 
However, what can be said is that the lower wind opacity in V3885 Sgr is
likely to make it the better testbed for probing links between the wind 
and the physics of the disk it emerges from.


\subsection{Time variability, and a comparison with the stronger UV
spectroscopic variables BZ~Cam and V603~Aql}
\label{subsec:discus2}

In section \ref{sec:timevar} we have presented trailed mean-subtracted
spectra of the N\Vl1240, Si\IVl1398 and
C\IVl1549~lines.  The time resolution of these series is 30\thinspace sec.
In one sense these time series are a disappointment because the broad
blueshifted absorption is revealed as being relatively steady in both
binaries.  There is none of the more obvious short term variability described 
in the UV spectra of V603~Aql (Prinja et al. 2000a) and BZ Cam
(Prinja et al. 2000b). A re-examination of IX Vel difference spectra at 10-sec 
time-resolution has revealed no variability unseen in the 30-sec time-binning. 
It would appear then, on the limited basis
of a sample of 4 objects, that (i) systems selected for examination at
higher time resolution, because they are already known to be significantly
variable on the longer timescale of tens of minutes, continue to reward
with variability on the tens of seconds timescale (BZ Cam, V603~Aql), (ii)
systems not known to be markedly variable on the longer timescale remain as
such on the shorter (IX Vel, V3885~Sgr).  A possible point of distinction
between these two pairs of objects is that the variable objects are also
mooted as very low orbital inclination binaries ($i < 40^{\circ}$).
This difference permits the hypothesis that the variability is due to the
line-of-sight sampling some fluctuation in the angling of the innermost
disk-wind streamlines.  This assumes that there is something like a central
conical cavity in the outflow as sketched by Shlosman \& Vitello (1993) and
by Knigge, Woods \& Drew (1995) and substantiated by the numerical models
published by e.g. PSD. By contrast, at higher inclination the line-of-sight
will be well and truly embedded in the body of the approaching outflow at all
times.  Clearly it would be of benefit to be more sure of these binaries'
orbital inclinations.

There is another very striking point of contrast to note between these two
pairs of objects.  In the UV spectra of IX~Vel and V3885~Sgr, there is very
little redshifted line emission partnering the broad blueshifted absorption
-- only C\IVl1549 typically shows any at a significant level.  This is
quite common among high state non-magnetic CVs.  Yet in BZ Cam and V603~Aql
there is very strong redshifted line emission.  This difference could be
said to run right against expectation if BZ~Cam and V603~Aql really are much
more nearly face-on binaries, given that strong line emission is generally
associated with eclipsing systems ($i\ga70\degr$).  In this context
it is helpful to remember that exercises in the simulation of wind-formed
UV line profiles (e.g. Drew (1987), Shlosman \& Vitello (1993), Knigge
et al.
(1995)) together with the discovery of orbital-phase linked line profile
variations have led to the recognition that a component of the detected UV line
emission need not have a wind origin (see discussion of this in Drew, 1997).  Given this, the implication of the stronger redshifted line emission
in BZ~Cam and V603~Aql is either that these binaries are `the same' as
IX~Vel and V3885~Sgr and somehow the line emission is occulted at
higher orbital inclinations or there is an intrinsic physical difference
between these two pairs of objects that expresses itself via the
observed strength of the UV line emission.

In both IX~Vel and
V3885~Sgr there is evidence that narrow absorption features, sometimes
seen superposed on broad blueshifted absorption, exhibit an
organised motion that may prove to follow that of the accreting WD.
At a mean blueshift of $\sim-900$ \kms, these features are present only in
the data showing the most pronounced P~Cygni behaviour.   In the I1 and V3
timeseries of the
mean-subtracted line profiles,  the red absorption edge close to line centre
approximately follows the motion of the narrow features -- as makes sense if
both are due to the wind and hence follow the binary motion of the white
dwarf.  If this interpretation is correct, and there is no azimuthal
dependence of the line-of-sight velocity of these features, the detected
velocity changes provide a constraint on the orbital phase observed
and the white-dwarf $K$-velocity.

It is likely that the manner in which these narrow absorption features
vary gives a clue as to their origin.  On noticing them in \iue\
high-resolution data on IX Vel, Mauche (1991) reported three categories of
explanation for similar features seen in O-star spectra as summarised by
Henrichs (1988). The first concept is that an absorption dip may be caused by
a plateau in the outflow velocity law which allows a larger absorbing column
to build up over a small change in line-of-sight velocity.  The second
is that an absorber density enhancement due to e.g. an ionization effect
may have the same consequence.  Lastly, one may appeal to non-monotonicity due
to e.g. shocks sweeping through the flow as the cause of narrow dips.  Only
the velocity law plateau concept warrants further examination in the context
of these disk winds.  This is because there are obvious problems with the
other two.  First, the density enhancement idea sits uneasily with the
observed presence of these dips in transitions sampling a wide range of
ionization states together with their absence from a subordinate line
(C\IIIl1175) that should be favoured at an enhanced density.
Secondly, the difficulty with the shock-swept, or otherwise unstable, disk
wind class of explanation is the persistence and near stationarity of the
dips at a velocity well below the terminal value. Table \ref{tab:iue_vels}
list those \iue\ datasets in which there are clearly narrow features superposed
on the C\IVl1549 line, along with the locations of these narrow
features (see also Mauche 1991, Prinja \& Rosen 1995).  The similarity 
between the dip blueshifts in IX~Vel and V3885~Sgr, together with their 
earlier appearance in \iue\ observations, offers the prospect of an origin 
rooted in some basic property common to both binaries.
\begin{table}
\caption{Location of narrow features found superimposed on the C\IVl1549 doublet in high resolution \iue\ data of IX Vel. Velocities are given in \kms.}
\label{tab:iue_vels}
\begin{tabular}{@{}lll}
\hline
\iue\ dataset	&C\IVl1548.2	&C\IVl1550.7	\\
\hline
SWP22353	&$-1730\pm70$,$-850\pm70$&$-960\pm60$		\\
SWP25449	&$-1000\pm90$		&$-980\pm80$		\\
SWP29618	&$-1040\pm70$		&$-1010\pm70$		\\
\end{tabular}
\end{table}

It should be mentioned that similar narrow absorption dips are seen in
association with O\VIl1035 blueshifted absorption in outburst
spectra of U~Gem \cite{2001froning}. These data are complementary to those
presented here in that the orbital phase coverage is close to complete, while
the spectral resolution and S/N ratio is reduced.  The greater duration of the
FUSE observations reveals that the O\VI\ dips vary about their mean blueshift
of $\sim 500$~km s$^{-1}$ by no more than a few tens of kilometres per
second.  The more marked variation is in equivalent
width -- the dips are seen to fade and reappear in a somewhat erratic
fashion.

In one epoch of observation of V3885~Sgr, V2, the mean-subtracted timeseries
reveals a quite subtle and gradual change of a different character: the
seeming weakening of the wind signature in e.g. the C\IV~line profile
during the second 1000\thinspace sec of the observation.  This might be
interpreted as a temporary calming of the outflow.  The change, at the same
time, in the appearance of the 1300\thinspace\AA\ absorption feature
(figure~\ref{fig:v3885_lines}) points to a different explanation.  This
typically broad shallow depression is thought of as originating in the disk
atmosphere.  The fact that it develops superposed narrow absorption dips at
close to the rest wavelengths of the Si\III~multiplet, the dominant
contributor, strongly hints at a change in the visibility of the inner disk
-- and along with it, the disk wind.  This is not exceptional behaviour,
in that similarly-narrow unshifted absorption dips are the norm for
U~Gem in outburst \cite{2001froning}, and have been observed at a range
of orbital phases in the UV spectra of UX~UMa and RW~Tri (Mason, Drew \&
Knigge 1997).  That the narrow absorption is fleeting in V3885~Sgr
may be a hint of a somewhat lower, but not too low, orbital inclination.

Finally, we do not find frequent signatures of wind inhomogeneity on
or near the dynamical timescale in our data for either IX~Vel or
V3885~Sgr - just one spell of variability, the $\sim350$\thinspace sec
`wobble' seen in the V3 dataset is all there is to report.  This
indicates that structure extending over a significant fraction of the
projected area of the UV continuum source is rare.  If the
inhomogeneity described by PSD exists, it must be finer in scale.  One
way of achieving this in a model may be to allow the flow to be fully
3-dimensional.  Specifically, the derived geometry and dynamics of a
3-D wind is likely to differ from those of the axisymmetric 2-D wind
modelled by PSD.  For example, the knots discovered by PSD correspond
to rings in 3-D with axisymmetry imposed, but in a full 3-D simulation
they may be replaced by spirals or even a foam of very small blobs.
If there really is no significant inhomogeneity, then the
radiation-driven wind models of PSD would demand that around a half or
more of the luminosity is radiated quasi-spherically, or that flow
co-rotation is enforced by e.g. a strong, large-scale, ordered
magnetic field threading the disk.  However, before such adjustments
are considered, there is a more basic aspect of the radiation-driven
wind model to review.


\subsection{Implications for the mass loss mechanism}
\label{subsec:discus3}

If radiation pressure powers the mass loss from IX~Vel and from
V3885~Sgr, there are good reasons to expect a strong scaling of the
rate of mass loss to bolometric luminosity.  Indeed, using the plot of
effective Eddington number ($\Gamma$, the ratio of bolometric
luminosity to the Eddington limit) against mass loss rate that was
given in Drew \& Proga (2000), it is possible to estimate how strong
this dependence should be.  On adopting the mass accretion rate and WD
parameters derived for IX~Vel by Beuermann \& Thomas (1990), we would
expect $\Gamma$ to be about $4\times10^{-4}$.  For V3885~Sgr, this
quantity is unlikely to be greater and is probably smaller.  This
places both systems in the regime where an exponential dependence is
expected.  Even if it were the case that $\Gamma$ were high enough to
yield a gentler dependence, this would still be at least as steep as
$\mdot_{\rm{w}} \propto \lbol^2$.  So, unless there is a significant
countermanding effect due to shifts in wind ionization, we should
expect to see a strong positive correlation between the strength of
wind features in the ultraviolet and the ultraviolet continuum level
-- given that the latter is expected to scale very nearly linearly
with mass accretion rate for fixed WD parameters.

The time-averaged spectra presented in section 3 do not sit well with
this expectation.  In V3885~Sgr the weakest wind signatures are seen
when the ultraviolet continuum is brightest (V2), and strongest when
the UV continuum is 20 to 30 per cent fainter (V3). In our IX Vel data
(I3), the wind all but disappears when the UV continuum is at a level
intermediate between the levels observed at the other two epochs.
This absence of trend is seen in a setting where the evidence from the
UV spectral lines is that ionization shifts are not significant and
abundant ion stages are represented (see section 3.1).  This gives us
confidence that there is a direct (if uncalibrated) mapping from the
mean strength of blueshifted absorption onto mass loss rate, from the
lines that we observe with \hst.

Caution must be exercised regarding the significance of IX Vel's
apparent UV flux changes because these data were obtained through an
aperture small enough that the STIS flux throughput could have been
different at each epoch.  However, it is very unlikely that we are
misled in the case of V3885~Sgr since the data were obtained through
the larger 0.2 arcsec$^2$ aperture that has been characterised as
yielding UV fluxes to with 10 per cent of their photometric values.

\begin{figure*}
\vspace*{12cm}
\includegraphics{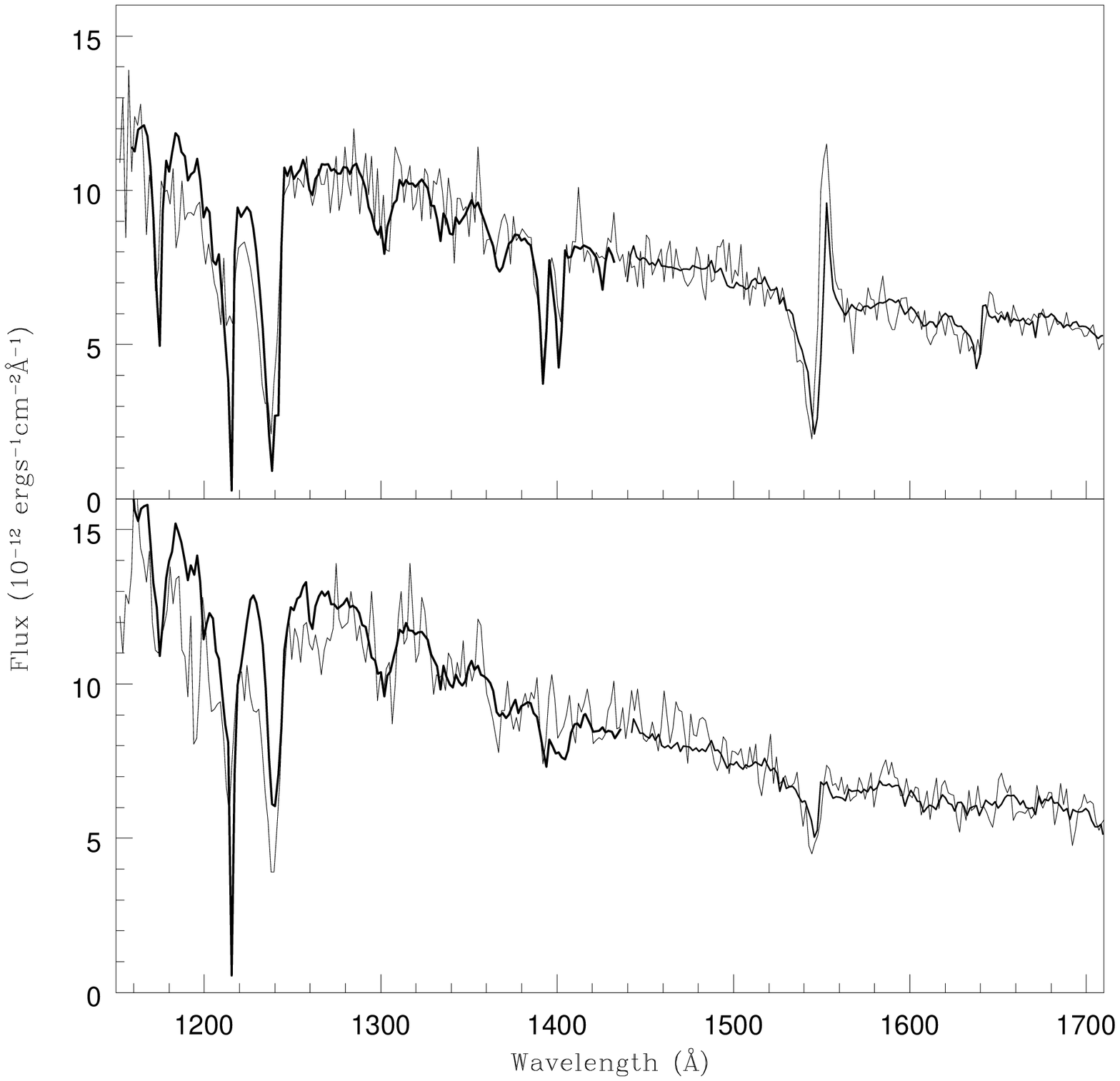}
\caption{I1 and I3 rebinned to $\Delta\lambda=1.8$\thinspace\AA\ and overplotted
on an \iue\ low-dispersion spectrum. Top frame - I1 plotted over data
set SWP19765 . Bottom frame - I3 plotted over data set SWP42994. The
\hst\ I1 and I3 spectra have both been multiplied by constants in order to 
rescale their flux levels to those of the \iue\ spectra with which they are
compared.}
\label{fig:ix_iue}
\end{figure*}

Since the failure to find a correlation between wind features and
continuum level is inconsistent with our expectation, it is important
to see whether independent evidece of the effect can be found in the
\iue\ database. We have done this for the case of IX Vel
as there is a particularly extensive set of \iue\ spectra to draw upon
(12 at high resolution and 28 at low resolution).  First of all, we note 
that the \iue\ recorded flux level in the 1260--1270\thinspace\AA\
range varies from $9.1\times 10^{-12}$ to $13.6 \times 10^{-12}$ 
erg s$^{-1}$ cm$^{-2}$\thinspace\AA$^{-1}$.  The \hst\ data at epochs I1 and I3
(table 3) are consistent with this, while I2 was a time of seemingly
unusually low continuum flux.  Given that the $V$ band flux is known
to vary secularly by up to $\sim$0.9 mags, the factor of 2 drop between I1 
and I2 is not so large it must be instrumental in origin.  Secondly, we have 
rebinned the weak-wind I3 spectrum to match the low resolution mode of
\iue\ in order to determine if such a state was also recorded by \iue.
We find that three of the 28 `LORES' spectra (SWPs 19765, 19766 and
26081) resemble the rebinned I3 data: to illustrate this, in figure 
\ref{fig:ix_iue} 
an overplot of the rebinning on SWP 19765 is presented.  If the I1 data
are similarly degraded in spectral resolution, these too can be matched
to examples of \iue\ spectra (see also figure \ref{fig:ix_iue}). These simple 
comparisons
allow the conclusion that the \hst\ observations describe the same
basic phenomenology as \iue\ (but more fully, of course).

\begin{figure}
\vspace{16cm}
\includegraphics{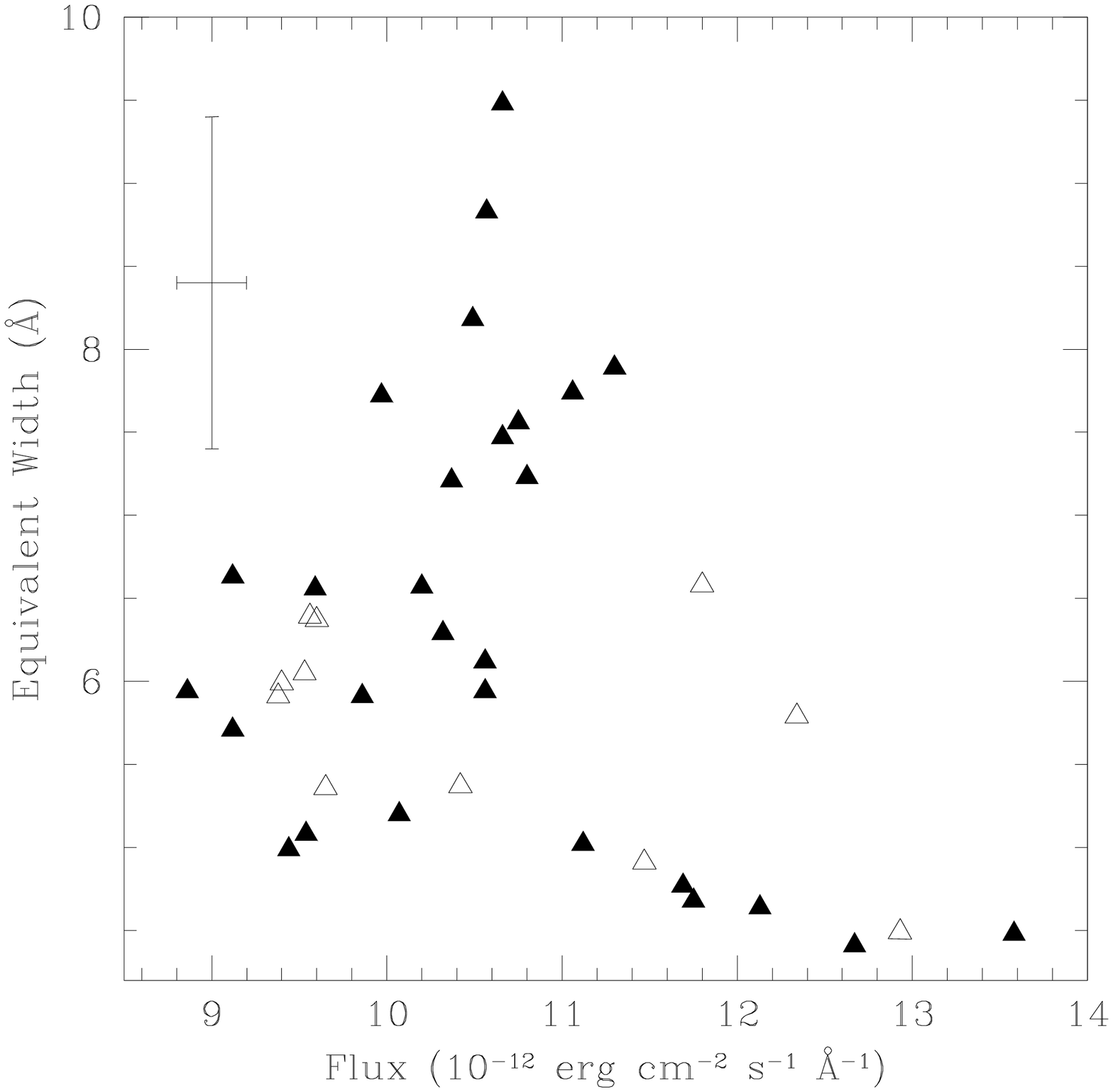}
\includegraphics{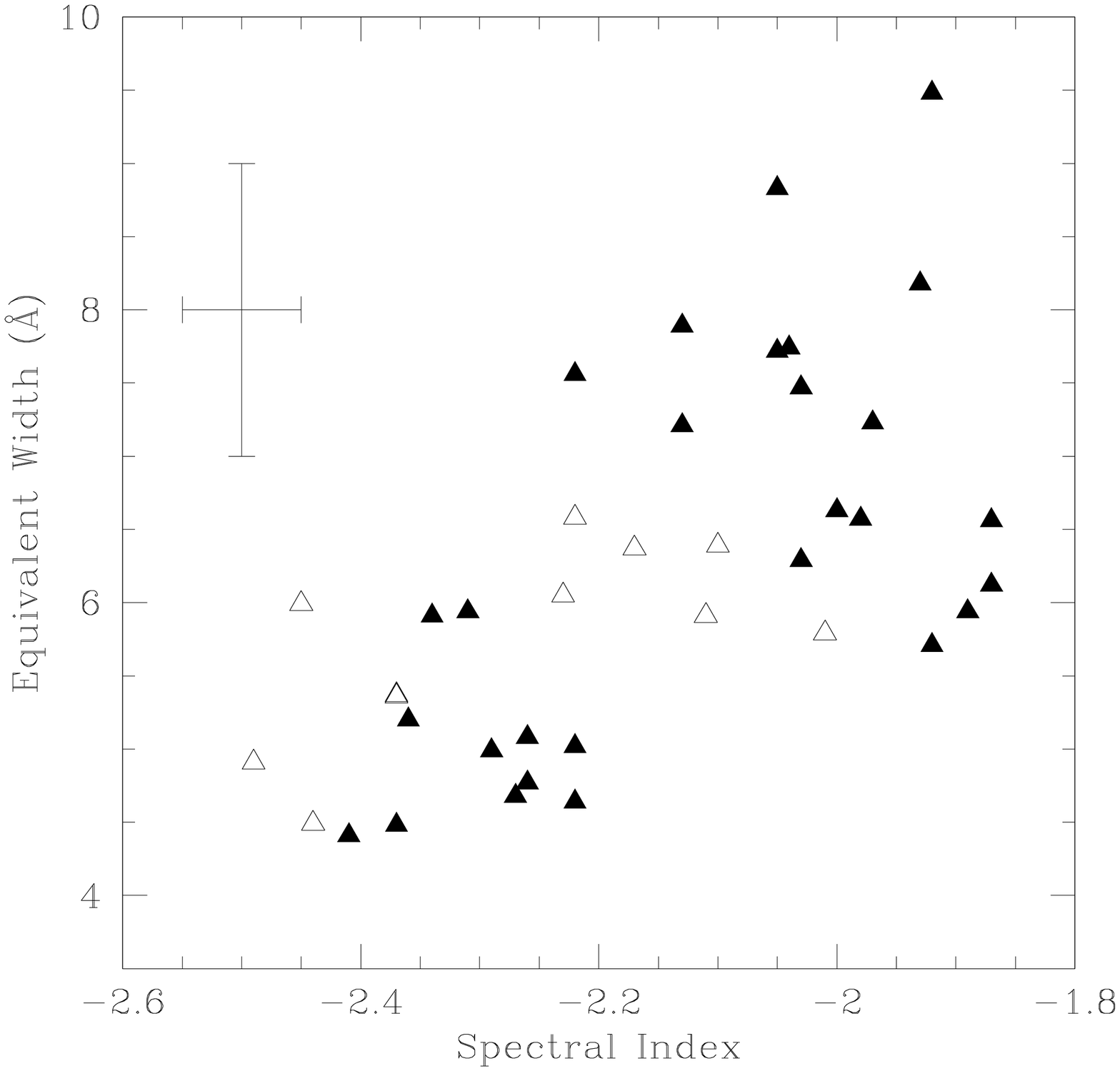}
\caption{Plot of C\IV$\lambda1549$ absorption equivalent width
against, a) continuum flux in the 1260--1270\thinspace\AA\ range (above) and b)
spectral index (below) for IX Vel \iue\ archive data. Solid triangles
represent LOWRES data and open triangles represent HIRES data. An
example of the estimated typical error bar is given in the top
left-hand corner of each plot. Pearson correlation coefficients for a
linear fit to these data are given in table \ref{tab:pearson}.}
\label{fig:ewflux}
\end{figure}


One of the admirable features of the final archive of \iue\ data is
its photometric consistency.  This allows us to carry out the
experiment on IX~Vel archive data of evaluating the strength of
correlation between wind activity, as signalled by the equivalent
width of C\IVl1549 blueshifted absorption, with (i) UV flux level,
again measured over the narrow band, 1260--1270\thinspace\AA, (ii) UV
continuum spectral index (fitting the continuum to a power law using
much the same method as adopted for the \hst\ summed spectra).  These
pairs of quantities as measured from \iue\ archive data are plotted in
figure \ref{fig:ewflux}.  In addition we have formally evaluated the
Pearson correlation coefficients between these variable pairings, and
also between the UV flux level and spectral index The results obtained
are shown in table \ref{tab:pearson} . In support of what we have seen
in the \hst\ data, there seems to be at best scatter, or perhaps even
a weak anti-correlation, between the C{\sc iv} absorption equivalent
width and UV brightness (figure
\ref{fig:ewflux}a).  By contrast, there does seem to be a correlation between 
the C\IV\ line and UV spectral index in the same sense as picked out 
earlier -- i.e. the slope of the spectrum reddens as the wind activity 
increases.  
\begin{table}
\caption{Pearson's correlation coefficient, r, for the relationship between 
continuum spectral index, flux and C\IV\ blueshifted absorption equivalent 
width in the \iue\ archive IX Vel data. $P(r)$ is the probability of a random 
distribution having a Pearson coefficient $\ge|r|$}
\label{tab:pearson}
\begin{tabular}{@{}lcc}
\hline
Variables		&$R$	&$P(r)$	\\
\hline
Spectral Index vs. Flux	&$-0.249$	&$0.127$\\
C\IV\ EW vs. Flux	&$-0.270$	&$0.096$\\
C\IV\ EW vs. Spectral Index	&$0.622$	&$2.36\times10^{-5}$\\
\end{tabular}
\end{table}

Circumstantially, if not absolutely conclusively, it would appear that
there is not the expected strong positive correlation between apparent
mass loss and luminosity.  A key prediction of the radiatively-driven
disk wind model is flouted.  The flow timescale in cataclysmic
binaries like IX~Vel and V3885~Sgr is likely to be a few tens of
seconds (assuming a Castor \& Lamers type velocity law from 40 to
5000\kms integrated over a distance $10\,R_{\rm{WD}}$). This is
sufficiently short that the wind dynamical configuration is able to
adjust continuously to the changing radiant flux as long as this does
not vary markedly on an even shorter timescale. In neither object is
there evidence of this either in our data or, historically, from
observations of flickering (e.g. Williams \& Hiltner 1984, Cowley et
al 1977).  Hence, it is not an option to invoke a concept of
e.g. hysteresis to explain the absence of the expected steep relation
between luminosity and mass loss rate.  We conclude that neither
IX~Vel nor V3885~Sgr knows they have to obey such a relation!

\subsection{Final Remarks}
\label{subsec:discus4}

Now that quantitative models of radiation-driven disk winds have
appeared, and there are high quality data to match them. it can be
seen more clearly that a factor other than line-driving is much more
likely to be decisive in powering these outflows.  Our findings add to
the doubts over the years that line-driving cannot sustain the mass
loss rates observation seems to imply (see Mauche \& Raymond 2000,
Drew \& Proga 2000).  The de facto exclusion of a class of models is
progress.  It is now time to examine the alternatives with renewed
intent.
 
MHD driving has long been viewed as a possible contributor or sole
cause of mass loss (Cannizzo
\& Pudritz 1988, see also comments in Drew 1991). This option has
become more attractive again in recent years as the magneto-rotational
instability has emerged as a favoured angular momentum transport
mechanism for CV disks (Balbus \& Hawley 1998, Tout 2000). Perhaps the
most promising alternative to a pure line-driven disk wind model is a
hybrid model where a wind is driven from the disk by some combination
of MHD and line forces.

Calculations combining the two driving mechanisms rigorously do not
exist yet. However we can refer to some schematic predictions of
radiation-driven disk wind models where the main role of an ordered
magnetic field is to impose angular velocity, rather than angular
momentum, conservation (Proga 2000).  For $L_{\rm{D}} \ga 0.001
L_{\rm{E}}$, imposing corotation on a line-driven disk wind model
increases the wind mass loss rate: the smaller the angle between the
poloidal streamlines and the disk midplane the larger mass loss rate
(cf. Blandford \& Payne 1982).  For $L_{\rm{D}}>$ a few times
$0.001L_{\rm{E}}$, $\mdot_{\rm{w}}$ is as in a pure line-driven disk
wind model. In terms of the $\mdot_{\rm{w}}$ vs $L_{\rm{D}}$
relationship, it means that for a relatively low disk luminosity
(i.e. $L_{\rm{D}} \ga 0.001 L_{\rm{E}}$), the mass loss rate can
increase or decrease depending on the streamline geometry required by
prevailing magnetic fields.  This control of $\mdot_{\rm{w}}$,
independent of the radiant luminosity, can potentially explain our
observations because these sample the luminosity range where this
additional dependence is expected.  As noted already, enforcement of
flow streamlines by a large scale magnetic field would also be
consistent with the near absence of detected wind inhomogeneity.

On the other hand, the fact that a radiation-driven wind can be
produced only when $L_{\rm{D}} \ga 0.001 L_{\rm{E}}$, has a basic
appeal in explaining the behaviour of other CVs such as dwarf novae
(DN).  The UV observations of DN through outburst obtained by \iue\
have provided evidence of mass loss signatures tracking luminosity.
For example, it has been noted during declines of DN from maximum
light that a decay of the UV continuum by a factor of $\ga 2$ can be
accompanied by a near disappearance of the blueshifted absorption so
prominent at maximum (Woods \& Drew 1990; Woods et al. 1992).  If this
correlation proves to be indirect rather than causal, we have work to
do to uncover the controlling physics behind it.  The way forward is
to identify for further development one of the many variants of MHD
disk wind model already explored in the literature that can provide a
convincing explanation, either with or without the assistance of
line-driving, of the continuing puzzle of these disk winds.  Once
again the accretion disk winds in CVs have emerged as particularly
open and rewarding laboratories for studying disk mass loss.  An
intriguing issue for the future is the existence or not of flow
collimation of CV disk winds if even CV disk winds are MHD powered
(see Knigge \& Livio 1998).  As these problems are tackled we can also
hope to obtain greater insight into the nature of disk winds in other
astrophysical settings.

\section*{Acknowledgements}
This paper is based on observations made with the NASA/ESA Hubble Space Telescope, obtained 
at the Space Telescope Science Institute, which is 
operated by the Association of Universities for Research in Astronomy, Inc.,
under 
NASA contract NAS 5-26555. These observations are associated with
proposal 8279.
The work presented in this paper was performed while Daniel Proga held
a National Research Council Research Associateship at NASA/GSFC.

\end{document}